\documentclass[prd,twocolumn,superscriptaddress,showpacs,nofootinbib,amsmath,amssymb]{revtex4}

\usepackage{bm}
\usepackage{amsfonts}
\usepackage{latexsym}
\usepackage[latin1]{inputenc}
\usepackage{graphicx}
\usepackage{amsmath}
\usepackage{rotating}
\usepackage{epsfig}

\def \nn  {\nonumber}

\newcommand{\crt}{\mathrm c}

\newcommand{\n}{\mathrm n}
\newcommand{\p}{\mathrm p}

\newcommand{\x}{\mathrm x}
\newcommand{\y}{\mathrm y}
\newcommand{\f}{\mathrm f}
\newcommand{\ar}{\mathrm r}

\newcommand{\al}{\alpha}

\def \nn  {\nonumber}

\def\jnl@style{}
\def\aaref@jnl#1{{\jnl@style#1}}

\def\aaref@jnl#1{{\jnl@style#1}}

\def\aj{\aaref@jnl{AJ}}                    
\def\apj{\aaref@jnl{Astrophys.~J.}}        
\def\apjl{\aaref@jnl{Astrophys.~J.~Lett.}} 
\def\apjs{\aaref@jnl{ApJS}}                
\def\apss{\aaref@jnl{Ap\&SS}}              
\def\aap{\aaref@jnl{Astron.~Astrophys.}}   
\def\aapr{\aaref@jnl{A\&A~Rev.}}           
\def\aaps{\aaref@jnl{A\&AS}}               
\def\mnras{\aaref@jnl{Mon. Not. R. Astron. Soc.}}             
\def\prd{\aaref@jnl{Phys.~Rev.~D}}        
\def\prl{\aaref@jnl{Phys.~Rev.~Lett.}}    
\def\qjras{\aaref@jnl{QJRAS}}             
\def\skytel{\aaref@jnl{S\&T}}             
\def\ssr{\aaref@jnl{Space~Sci.~Rev.}}     
\def\zap{\aaref@jnl{ZAp}}                 
\def\nat{\aaref@jnl{Nature}}              
\def\aplett{\aaref@jnl{Astrophys.~Lett.}} 
\def\apspr{\aaref@jnl{Astrophys.~Space~Phys.~Res.}} 
\def\physrep{\aaref@jnl{Phys.~Rep.}}      
\def\physscr{\aaref@jnl{Phys.~Scr}}       

\begin{document}


\title{The evolution of the $f$-mode instability in neutron stars and gravitational wave detectability}

\author{A. Passamonti}
\affiliation{Theoretical Astrophysics, IAAT, Eberhard Karls University of T\"{u}bingen, T\"{u}bingen 72076, Germany} 
\affiliation{INAF - Osservatorio Astronomico di Roma, Via Frascati 33, 00044 Rome, Italy}
\author{E. Gaertig} 
\affiliation{Theoretical Astrophysics, IAAT, Eberhard Karls University of T\"{u}bingen, T\"{u}bingen 72076, Germany}
\author{K.~D. Kokkotas}
\affiliation{Theoretical Astrophysics, IAAT, Eberhard Karls University of T\"{u}bingen, T\"{u}bingen 72076, Germany}
\author{D. Doneva}
\affiliation{Theoretical Astrophysics, IAAT, Eberhard Karls University of T\"{u}bingen, T\"{u}bingen 72076, Germany} 
\affiliation{Institute for Nuclear Research and Nuclear Energy, Bulgarian Academy of Sciences, Sofia, Bulgaria}

\date{\today}

\begin{abstract}

We study the dynamical evolution of the gravitational-wave driven instability of the $f$-mode in rapidly rotating relativistic stars.  
With an approach based on linear perturbation theory we describe the evolution of the mode amplitude and  
follow the trajectory of a newborn neutron star through  its instability window. The influence on the $f$-mode instability 
of the magnetic field and the presence of an unstable $r$-mode is also considered.
Two different configurations are studied in more detail; an $N=1$ polytrope with a typical mass and radius and a more massive polytropic $N=0.62$ model  
with gravitational mass $M=1.98 M_{\odot}$. 
We study several evolutions with different initial rotation rates and temperature and determine the gravitational waves radiated during the instability. 
 In more massive models, an unstable $f$-mode with a saturation energy of about  $10^{-6} M_{\odot} c^2$ may generate a gravitational-wave 
 signal which can be detected by the Advanced LIGO/Virgo detector from the Virgo cluster.
 The  magnetic field  affects the evolution and then the detectability of the gravitational radiation when its strength is 
higher than $10^{12}$~G, while the effects of an unstable $r$-mode become dominant when this mode reaches the maximum saturation value 
allowed by non-linear mode couplings. 
However, the relative saturation amplitude of the $f$- and $r$-modes  must be known more accurately 
in order to provide a definitive answer to this issue.  
From the thermal evolution  we find  also that the heat generated by shear viscosity during the saturation phase completely balances the 
neutrinos' cooling and prevents the star from entering the regime of mutual friction. The evolution time of the instability is therefore longer 
and the star loses significantly larger amounts of angular momentum via gravitational waves.

\end{abstract}


\maketitle

%
\section{Introduction}
\label{Sec:intro}
%

After a core-collapse supernova explosion, a nascent neutron star, if rapidly rotating,  may  develop non-axisymmetric instabilities and 
radiate a significant amount of gravitational waves~\cite{2009CQGra..26f3001O,2011GReGr..43..409A}. 
These instabilities may originate from different physical processes and grow on dynamical and secular timescales.

The dynamical instabilities in uniformly rotating stars occur at high rotation rates, the typical case of the bar 
mode instability, i.e. driven by the $l=m=2 f$-mode,  
 sets in   when the kinetic to gravitational potential ratio is   $T / |W| \gtrsim 0.255$~\cite{2007PhRvD..75d4023B, 2007CQGra..24S.171M}.
In stars with a high degree of differential rotation, a dynamical instability may develop even at considerably lower rotation rates, $T/|W| \gtrsim 0.01$, 
(see \cite{2009CQGra..26f3001O} and reference therein). 
Secular instabilities are driven instead by dissipative processes and thus develop on longer timescales but at smaller 
rotation rates. For instance, the viscosity driven bar mode instability  typically sets in when  $T / |W| \gtrsim 0.14$~\cite{2003CQGra..20R.105A}, 
while the gravitational-wave driven $l=m=2$ $f$-mode instability requires 
roughly $T / |W| \gtrsim 0.13$~\cite{1983PhRvL..51...11F}, and in more compact stars $T / |W| \gtrsim 0.07$~\cite{2010PhRvD..81h4055Z}. 
The conditions are much better 
for higher multipole $f$-modes which can be gravitationally unstable at even 
 lower rotation rates~\cite{1983PhRvL..51...11F}.

In rotating neutron stars a non-axisymmetric mode may be driven unstable by gravitational radiation 
via  the well-known Chandrasekhar-Friedman-Schutz (CFS) mechanism~\cite{1970PhRvL..24..611C,1975ApJ...199L.157F,1978ApJ...221L..99F}. 
This instability occurs when a mode  which is counter-rotating  with respect to the star  is 
seen as co-rotating by an inertial observer. The angular momentum radiated by gravitational waves induces an increasingly 
negative angular momentum of the mode which thus becomes unstable~\cite{1970PhRvL..24..611C,1975ApJ...199L.157F,1978ApJ...221L..99F}.
The mode's growth ends when some dissipative process suppress the instability.  

 Non-axisymmetric modes may be excited during the proto-neutron star formation, after the core bounce, and become CFS unstable 
as the star cools down below a critical temperature which depends on the star's model and rotation.  
The mode's amplification expected during the instability may generate a significant gravitational wave signal 
which can be potentially observed by the current and next generation of Earth based laser interferometers. 
When the gravitational waves will be observed,  the identification of the oscillation modes from the spectrum 
will help us to unveil the properties of the dense matter at super-nuclear densities and clarify the neutron star physics. 
The analysis of the observed Quasi Periodic Oscillations in Magnetars  already provided important 
insights~\cite{2007MNRAS.374..256S,2008MNRAS.385.2161S,2009PhRvL.103r1101S, 2011MNRAS.414.3014C, 2012arXiv1208.6443G}, and 
even more relevant results are expected when Astereoseismology will be applied to the gravitational-wave 
signal~\cite{Andersson:1996ak,1998MNRAS.299.1059A}.

Among the several modes which can be 
driven unstable, the $f$- and $r$-modes are the most important due to their relatively short growth timescale and an 
efficient emission of gravitational waves. The $r$-mode instability has attracted more attention so far as it is CFS  unstable 
at any rotation rate and may grow in about few tens of seconds 
in rapidly rotating stars.  There is in fact an extensive literature which studied 
the $r$-mode instability for various stellar models and considered it as a possible candidate 
for  limiting the star's rotation below the Kepler velocity (see for instance~\cite{2001IJMPD..10..381A, 2012ApJ...746....9P} and references therein). 
The growth of the $r$-mode can be however strongly limited by non-linear mode coupling with other inertial 
modes~\cite{2003ApJ...591.1129A, 2007PhRvD..76f4019B}.

The $f$-mode instability  occurs instead in rapidly rotating stars and preferable in more  compact objects  $(M/R > 0.2 )$~\cite{2011PhRvL.107j1102G}.
In Newtonian stellar models with stiff polytropic equation of state (EoS),  i.e. with polytropic index $N<1$, the $l=m=2$ $f$-mode is unstable  close to the maximum rotation rate, while 
in softer equations of state  ($N \geq 1$)  this mode is always stable~\cite{1990ApJ...355..226I, 1991ApJ...373..213I, 1995ApJ...438..265L}. 
The  multipoles for which the $f$-mode becomes unstable for smaller rotation rates 
are  the $ l =m = 3$ and 4, which have a large instability window and 
still a reasonable short growth time~\cite{1983PhRvL..51...11F,1990ApJ...355..226I, 1991ApJ...373..213I, 1995ApJ...438..265L}. 
The conditions are more promising in relativistic stellar models as the $f$-mode is driven unstable at smaller rotation rates,  
and also the quadrupole $f$-mode might have a significant  instability window~\cite{1998ApJ...492..301S, 2010PhRvD..81h4055Z, 2011PhRvL.107j1102G}. 

To date the evolution of the $f$-mode instability has been  studied only for the $l=m=2$ case 
 with non-linear dynamical simulations in Newtonian gravity. These works used 
both ellipsoidal configurations~\cite{1977ApJ...213..193D, 1995ApJ...442..259L} 
and compressible stellar models with uniform~\cite{2004ApJ...617..490O} and differential \cite{2004PhRvD..70h4022S} rotation. 
In these non-linear simulations, the effects of viscosity has been neglected and the 
bar mode instability is driven by a gravitational radiation reaction term which  has been incorporated in the Newtonian dynamical equations.  
In particular, the strength of this Post-Newtonian term has been artificially increased to shorten the instability evolution and 
make feasible its analysis within the simulation time. 
 For typical neutron stars parameter these non-linear studies find that 
 the gravitational-wave signal emitted during the bar mode instability may be detected 
 by Advanced LIGO from  a source located in the Virgo cluster. More importantly, 
   while the star  spins down 
 the mode's frequency decreases toward the more sensitive frequency band ($100~\textrm{Hz}$) of the gravitational-wave detectors.  
 
In the present work we study the evolution of the gravitational-wave driven $f$-mode instability with a different approach.  
It is based on the linear perturbation framework developed  for the $r$-mode instability in~\cite{1998PhRvD..58h4020O, 2002MNRAS.337.1224A}. 
From the evolution equations for the  mode energy, total angular momentum and temperature we  derive a system of ordinary differential equations to describe 
 the mode's amplitude, the star's rotation and the thermal evolution.  In this analysis we include the effects of viscosity, 
 magnetic field, and consider also the impact of an unstable $r$-mode on the $f$-mode instability. 
The coefficients of these evolution equations depend on the stellar model and mode properties. More specifically, for a given 
neutron star model we have to determine the $f$-mode frequency and eigenfunctions, and with these information calculate  
the relevant viscous and gravitational radiation timescales. 

 We consider rapidly rotating and relativistic models with uniform rotation and extract the $f$-mode properties (frequency and eigenfunctions) 
  from the time evolutions of the linearised dynamical equations. In particular, we simplify the problem by using the Cowling approximation, 
  i.e. we neglect the space-time perturbations in our linearised problem. 
  The accuracy of this approximation is to better than 20\% for the quadrupole $f$-mode, but decreases considerably for higher 
 multipoles~\cite{lrr-2003-3}.
  In a second step,  the mode's frequency and eigenfunctions are inserted in appropriate volume integrals which determine
   the viscous damping times and the gravitational radiation growth time~\cite{1991ApJ...373..213I}. 
 
\begin{figure*}
\begin{center}
\includegraphics[height=75.3mm]{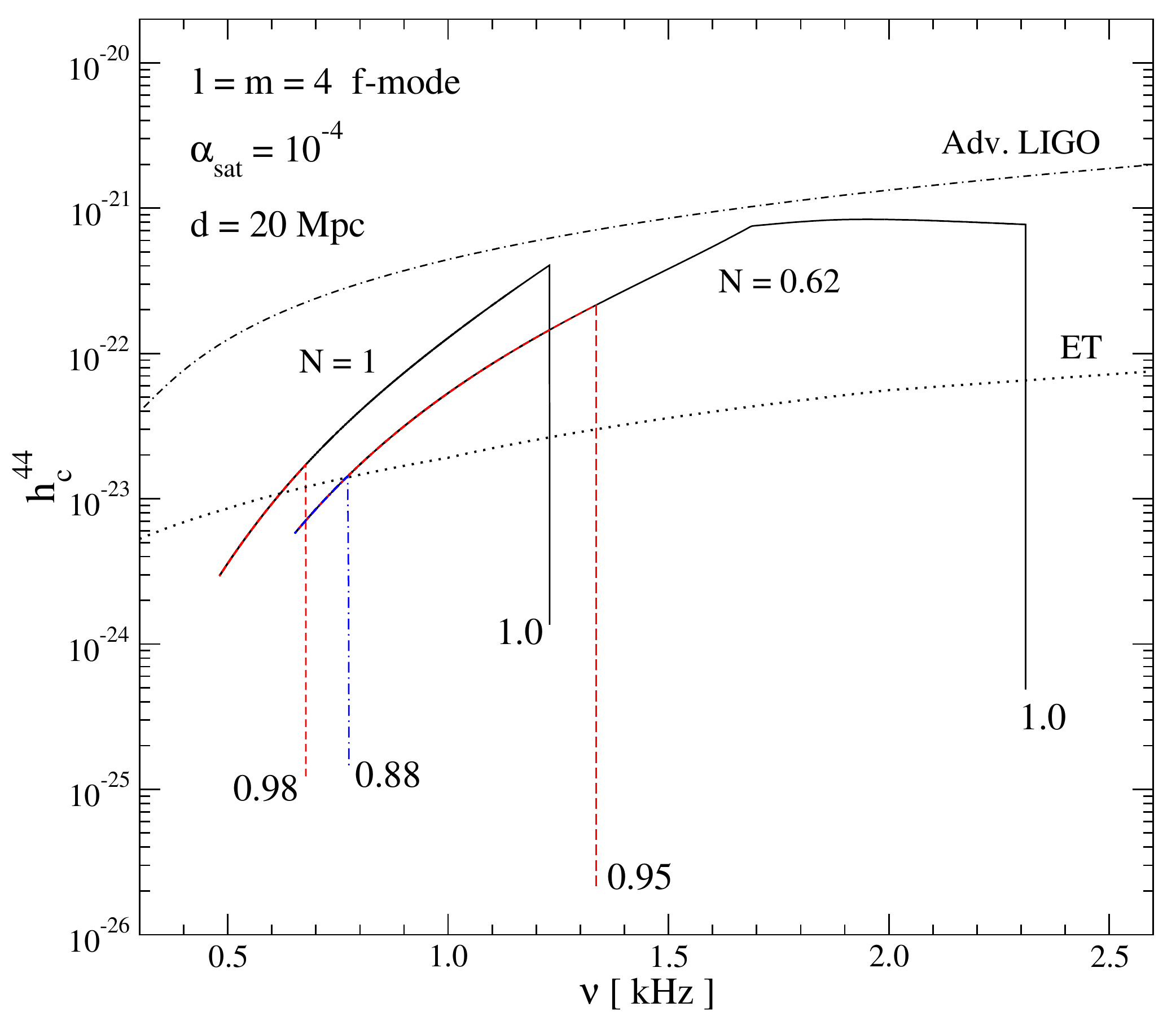} 
\includegraphics[height=75mm]{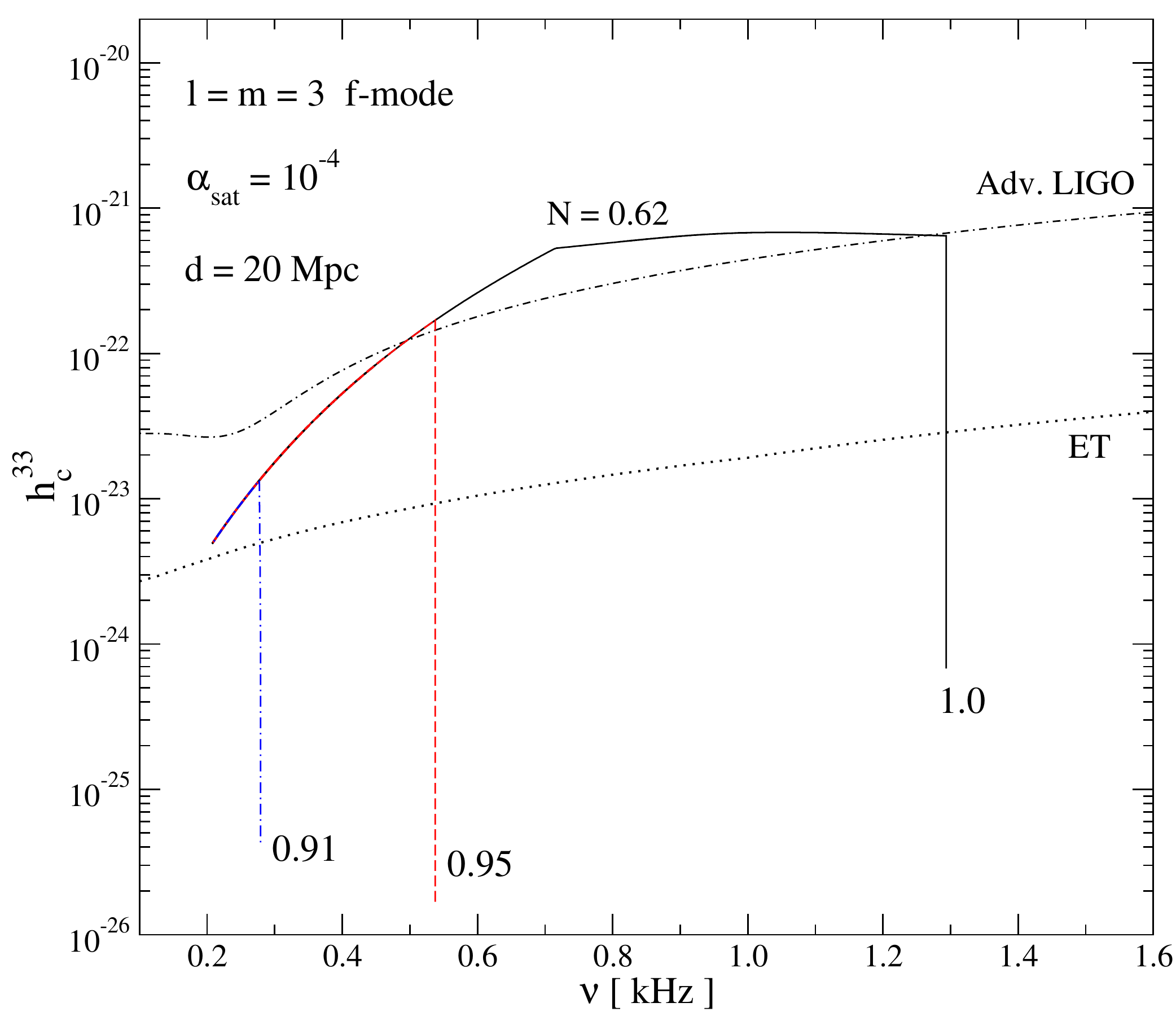} 
\caption{ Characteristic strain generated by the $f$-mode instability for  
a polytrope with $B_{\rm p }=10^{11}$G. The source is located at 20 Mpc and the saturation amplitude of the 
$f$-mode is set to $\alpha_{sat} = 10^{-4}$ (see Sections~\ref{sec:modevol} and~\ref{sec:FM}). {\it Left panel:} The gravitational-wave signal emitted by the $l=m=4$ $f$-mode 
for the two stellar models with $N=1$ and $N=0.62$.  The number near
the vertical lines denotes the initial rotation rate $\Omega /
\Omega_{\rm K}$. 
{\it Right panel:} The signal emitted the by $l=m=3$ $f$-mode for the more massive model with $N=0.62$. The sensitivity curves of  Advanced  LIGO and the Einstein Telescope (ET)~\cite{LIGO, ETel} are shown in both panels.
The gravitational-wave signal generated during the $f$-mode instability may be detected by ET for the most part of the instability 
window of the $N=0.62$ model. In particular, the $l=m=3$ $f$-mode may be detectable even by Advanced LIGO for $\Omega \geq 0.95 \Omega_{\rm K}$.
\label{fig0}}
\end{center}
\end{figure*}

  We study in more detail two polytropic neutron star models with different compactness and maximum rotation limit. 
  The first is a star with polytropic index $N=1$ and gravitational mass $M = 1.4  M_{\odot}$, 
   while the second is an $N=0.62$ polytrope with gravitational mass $M =1.98  M_{\odot}$.   
   Considering several configurations we find that the gravitational-wave signal generated by an unstable $f$-mode 
   may be potentially detected with the Einstein Telescope (ET) from a source located in the Virgo cluster.  
   For the more massive stellar model the signal may be even detected  by the Advanced LIGO/Virgo detectors.  
    For instance, the gravitational characteristic strain generated by the $l=m=3$ and 4 $f$-mode is shown in Figure~\ref{fig0} for the $N=1$ and $N=0.62$ polytropic models with a relatively 
    weak magnetic field, $B_{\rm p}=10^{11}~\textrm{G}$. This characteristic strain is determined by integrating in time the signal generated by 
    an $f$-mode during the instability and setting a maximum saturation energy of $E\sim 10^{-6} M_{\odot} c^2$.  
    However  in order to not overestimate the signal  when the instability evolves on long timescales, we have considered 
    a maximum detector integration time of 1 year.  
    For more details on these results and the effects of higher magnetic fields and $r$-modes see Sec.~\ref{sec:GW}.

  A general expectation is that superfluidity should further restrict the parameter space of the instability.   
  In fact, if a star cools down below the transition temperature where the neutrons of the core become superfluid,  
  the $f$-mode instability should be efficiently suppressed by the mutual friction force~\cite{1995ApJ...444..804L}. However,  for typical   stellar parameters 
   we find that 
  the heat generated by shear viscosity during the saturation phase 
  of an unstable $f$-mode may counterbalance the neutrino cooling.  In fact, our results show that an unstable star may follows a quasi isothermal trajectory 
  within the instability window without cooling below the critical  temperature of the core's neutrons. As a result the evolution time is longer
  and the star loses more angular momentum via the emission of gravitational waves.

  We present the formalism for studying the $f$-mode instability in 
  Sections~\ref{sec:eq} and~\ref{sec:evol}, while in the Appendix we provide 
  the equations for the gravitational-wave instability driven simultaneously by two modes. 
  The results for the instability evolution are described  in Section~\ref{sec:res}, and the gravitational-wave signal 
  is determined in Section~\ref{sec:GW}. The concluding remarks can be found in Section~\ref{sec:conc}.

\section{The $f$-mode instability} \label{sec:eq}

The standard approach for studying secular instabilities is based on linear perturbation theory and consists 
of two distinct steps. First,  we determine the main properties of a mode (frequencies and eigenfunctions), which will be inserted in a second step 
 into appropriate volume integrals for the estimation of the relevant timescales due to  viscosity and gravitational radiation. 
Since these timescales are generally much longer than the oscillation period of the mode, their properties can be determined 
by solving the inviscid problem~\cite{1991ApJ...373..213I}.  

In this work, we consider relativistic models of rapidly rotating neutron stars and 
determine the $f$-mode frequency and eigenfunctions by post-processing time evolution data of the relativistic perturbation equations. 
All the details of the formalism and numerical methods for studying the oscillations of rapidly rotating stars 
is given in~\cite{2008PhRvD..78f4063G, 2011PhRvD..83f4031G}.

In a realistic star, an oscillation mode approximately evolves as $e^{i \omega t - t/\tau }$,  
where $\omega$ is the mode's  frequency and $\tau$ the damping/growth timescale. 
A way to determine $\tau$ is to calculate the variation in time of the mode's energy given in the rotating frame by 
\begin{equation}
E  = \frac{1}{2} \int dV \left[  \rho \, \delta u^{i} \delta u_{i}^{*} + \frac{1}{2} 
\left(  \delta \rho \, \delta h^{*} + \delta \rho^{*} \delta h   \right)  \right] 
\label{eq:En}  \,  , 
\end{equation}
where the asterisk denotes a complex conjugate and the integral extends over the volume of the star. 
The quantity $\rho$ represents the background mass density, while  $\delta \rho$, 
$\delta h$ and $\delta u^{i}$ are perturbations of the mass density, the enthalpy and the velocity, respectively. 
The latin indices  denote the spatial components of a generic four vector.  Equation~(\ref{eq:En})  
is also defined as the canonical energy  in the rotating frame~\cite{1978ApJ...221..937F}.

As long as the energy of a mode depends quadratically on its perturbations, one can show that~\citep{1991ApJ...373..213I}
\begin{equation}
\frac{ d E}{dt} = - \frac{2 E}{ \tau} \label{eq:dEdt}  \,  , 
\end{equation}
where the global timescale $\tau$ of equation~(\ref{eq:dEdt}) results from the individual dissipative mechanisms that act on the mode. 
For stars in which only gravitational waves, shear and bulk viscosity dissipate energy, we have~\cite{1991ApJ...373..213I}:
\begin{equation}
\frac{1}{\tau} = \frac{1}{\tau_{\rm gw}} + \frac{1}{\tau_{\rm s}} + \frac{1}{\tau_{\rm b}}  \, .
\label{eq:tauT} 
\end{equation}

The gravitational radiation timescale can be calculated by using the standard multipole formula\footnote{In our calculations 
we correct the $\tau_{\rm gw}$ determined via equation~(\ref{eq:taugw}) with a constant factor equal to 3.  This procedure provides a better agreement with the 
relativistic results (no Cowling approximation) of the gravitational radiation timescale; see Doneva $\&$ Kokkotas (in preparation) for more details.}~\cite{1980RvMP...52..299T}, 
\begin{align}
& \frac{1}{\tau_{\rm gw}} =  \frac{\omega}{2E}  \sum_{l\ge2} N_{l} \left( \omega - m \Omega \right) ^{2 l+1} \left( \left| \delta D_{lm}  
 \right|^2 + \left| \delta J_{lm}   \right|^2\right) ,   \label{eq:taugw}
\end{align}
where $\omega$ is the mode frequency as measured in the rotating frame, $\delta D_{lm} $, $\delta J_{lm}$ are the 
mass and current multipole moments and $N_l$ is given by
\begin{equation}
N_l = \frac{ 4\pi G }{c^{2l+1}}  \frac{ \left(l+1\right) \left(l+2\right)  }
{ l \left( l -1\right) \left[ \left( 2 l + 1 \right) !! \right] ^2  } \,  ,
\end{equation}
where $!!$ denotes a double factorial.

 Since the $f$-mode mainly radiates via the mass multipole moments which are determined by
\begin{equation}
\delta D_{lm} = \int dV r^l \delta \rho \, Y_{lm}^{*} \, ,  \label{eq:Dlm} 
\end{equation}
we neglect the current multipole moments for computing its damping times. 

The dissipative timescales due to bulk- and shear-viscosity can be determined by 
the following volume integrals
\begin{align}
& \frac{1}{\tau_{\rm b}}  = \frac{1}{2E}   \int dV  \zeta \, \delta \sigma  \delta \sigma^{*} \, , \label{eq:tb}  \\
& \frac{1}{\tau_{\rm s}}  =  \frac{1}{E}  \int dV \eta \, \delta \sigma^{ij} \delta \sigma_{ij}^{*}  \label{eq:ts} \, ,
\end{align}
where  $\zeta$ and $\eta$ are the bulk and shear viscosity coefficients, while the 
stress tensor $\delta \sigma^{ij}$ is given in terms of velocity perturbations, 
\begin{align}
 \delta \sigma^{ij} & = \frac{1}{2} \left(   \nabla^{i} \delta u^{j} +  \nabla^{j} \delta u^{i}    
- \frac{2}{3} g^{ij} \nabla \delta \sigma  \right) \, , \\ 
\delta \sigma  & =  \nabla_j \delta u^j  \, ,
\label{visc2}
\end{align}
and $g_{ij} $ is the spatial metric tensor.  

The bulk and shear viscosity coefficients depend on the state and composition of 
the neutron star matter. For a neutron, proton, electron (npe) matter in normal state, i.e.\ with no superfluid/superconducting components, these coefficients  
can be written in a simple analytical form~\cite{1979ApJ...230..847F,1991ApJ...373..213I}.
More specifically, the bulk viscosity coefficient is given by~\cite{1989PhRvD..39.3804S}
\begin{equation}
\zeta = 6 \times 10^{-59}   \rho^2 \,  \omega^{-2} \, T^6 \,  \textrm{g cm}^{-1} \textrm{s}^{-1} \,  ,   
\label{eq:saw} 
\end{equation}
where $T$ is the star's temperature.
This expression is strictly valid for ``small'' oscillation amplitudes, within  the so-called sub-thermal regime. 
If the mode amplitude is sufficiently large, non-linear terms may become important  and will
increase the strength of the bulk viscosity~\cite{2012PhRvD..85b4007A,2012PhRvD..85d4051A}. 
However, the impact of the non-linear bulk viscosity on the $f$-mode instability is moderate and 
appears at rather large mode amplitudes~\cite{2012MNRAS.422.3327P}. 
In the present work we limit the $f$-mode growth to comparatively small amplitudes and then neglect the effects of non-linear  bulk viscosity. 

In normal npe matter, shear viscosity is dominated by neutron collisions 
and can be parametrized with the following coefficient~\cite{1979ApJ...230..847F,1991ApJ...373..213I}:
\begin{equation}
\eta = 347  \rho^{9/4}  \, T^{-2} \,  \textrm{g cm}^{-1} \textrm{s}^{-1}  \, .  \label{eq:eta}
\end{equation}

For an inviscid star, there is now the following situation: if the star is oscillating at a sufficiently high rotation rate, the condition for the secular CFS-instability, $\omega \left( \omega - m \Omega \right) \le 0$, might be fulfilled. In this case, the stellar rotation drags a counter-rotating mode into corotation as seen from an inertial observer, leading to a gravitational wave driven, exponential grow of the initial perturbation (it is $\tau_{ \rm gw} \leq 0 $ then; see equation~(\ref{eq:taugw})). However, the other dissipative mechanisms described above also operate in realistic stars and tend to stabilise the mode evolution. The instability onset and the mode growth is therefore controlled by the global timescale which has to be $\tau \leq 0$. 
This global timescale  is generally a function of the star's rotation rate $\Omega$ and temperature $T$ so that the instability has a natural representation in a $T-\Omega$ plane, where the region above the critical curve for $\tau = 0$ denotes the so-called instability window (e.g. see Figure~\ref{fig1}). 

This description of the instability is clearly static as no information about the evolution of the star and mode amplitude is provided. 
The evolution problem  will be addressed in the next Section.

\section{Evolution equations for the $f$-mode instability} \label{sec:evol}

To study the evolution of the $f$-mode instability we derive a system of  equations  
by using and extending the formalism already developed for the $r$-mode instability~\cite{1998PhRvD..58h4020O}. 
Our equations consider also the effects of thermal heating, magnetic torque and the impact of an  
unstable $r$-mode on the evolution of the $f$-mode instability and its gravitational-wave signal. 
The formalism  we develop here is general, it will be used in Sec.~\ref{sec:res}  to study the gravitational-wave instability 
of the $l=m=3$ and $4$ $f$-modes.

In the non-linear study of the secular bar-mode instability~\cite{1977ApJ...213..193D, 1995ApJ...442..259L, 2004ApJ...617..490O,2004PhRvD..70h4022S}, 
it is  reasonable to discern two dynamical phases of the $f$-mode instability, namely the mode's exponential growth 
and its non-linear saturation~\cite{1998PhRvD..58h4020O}. As the star enters the instability window, the mode  grows exponentially 
while the star slowly spins down on viscous 
timescales. This initial growth phase ends either when non-linear dynamics or dissipative processes 
saturate the mode amplitude at a finite value. After this point, the mode amplitude 
is nearly constant while the star loses angular momentum via 
 gravitational waves~\cite{2004ApJ...617..490O, 2004PhRvD..70h4022S}. 
  Eventually, the star leaves the instability window and the f-mode is damped by gravitational radiation.

The dynamical equations that describe these two regimes  of the instability  can be derived using the relations for the canonical mode energy (in the rotating frame) and the stellar angular momentum.
 In this section we focus on a star which is driven unstable by a single $f$-mode. In the Appendix we 
generalise the equations for studying the case in which two modes can be simultaneously driven unstable by gravitational radiation, 
and we apply later this formalism for the case of a simultaneous f- and r-mode instability.

\subsection{Mode evolution and star's spin-down} \label{sec:modevol}

The mode energy can be considered as a function of the 
mode amplitude and the stellar rotation rate.  Hence, we may  assume that 
$E = \alpha \tilde E\left( \Omega \right)$, where the parameter  $\alpha$ denotes the mode amplitude and $ \tilde E\left( \Omega \right)$ describes the angular velocity dependence. 
It is worth noticing that in contrast to~\cite{1998PhRvD..58h4020O, 2012MNRAS.422.3327P} we define $\alpha$ from the mode energy, 
hence  the fluid perturbations scales as $\sim \alpha^{1/2}$. 
After taking the time derivative,  equation~(\ref{eq:dEdt}) can be written as 
\begin{equation}
\frac{d \alpha}{dt} + \alpha \frac{d \ln \tilde{E}}{d\Omega} \frac{d\Omega}{dt} = -   \frac{2 \alpha}{\tau} \, . 
\label{eq:evol1}
\end{equation}

A second equation can be derived from the evolution of the angular momentum,  
\begin{equation}
\frac{dJ}{dt} = \frac{dJ_{gw}}{dt} + \frac{dJ_{mag}}{dt} \label{eq:dJdt} \,  ,
\end{equation}
where $J = J_s + J_c$ is the total angular momentum, which consists of the star's angular momentum $J_s$ and 
the canonical angular momentum of the mode $J_c$. The latter one can be related to the mode energy by the well known 
equation~\cite{1978ApJ...221..937F}:
\begin{equation}
J_{c} = - \frac{m }{\omega}  E  \,  .  \label{eq:Jc}
\end{equation}
The dissipative terms in equation~(\ref{eq:dJdt}) are the 
gravitational radiation torque, 
\begin{equation}
 \frac{d J_{gw}}{ dt} = - \frac{2  J_c}{ \tau_{ \rm gw}}  
\end{equation}
and the magnetic torque $d J_{mag}/dt$. 

We consider a standard oblique rotator model in vacuum with a dipolar magnetic field, which has~\cite{1983bhwd.book.....S}
\begin{equation}
\frac{d J_{mag}}{dt} = - \frac{ \sin ^2 \chi }{6 c^3} B_p^2 R^6 \Omega^3 \, , \label{eq:dJmagdt}
\end{equation}
where $R$ is the stellar radius, $B_{\rm p}$ is the magnetic field at the magnetic pole and $\chi$ is the inclination angle 
between the rotation and magnetic axes.  For simplicity we assume an orthogonal rotator ($\chi=\pi/2$). 
This configuration  provides also a magnetic spin-down comparable to that of a standard pulsar model 
 with magnetosphere~\cite{1983bhwd.book.....S,2006ApJ...648L..51S}.

From equation~(\ref{eq:Jc}) and the functional form of $E$ we can write the canonical angular momentum  
 as $J_c = \alpha \tilde J_c\left( \Omega \right)$. In this way we can simplify the calculation and determine 
from equation~(\ref{eq:dJdt}) the following expression:
\begin{equation}
\frac{d \alpha}{dt} +  \frac{1}{\tilde J_c }  \left( \frac{dJ_s}{d\Omega} + \alpha \frac{dJ_c}{d\Omega} \right) \frac{d\Omega}{dt} 
= - 2   \left( \frac{\alpha}{ \tau_{ \rm gw}} + \frac{1}{\tau_{ \rm mag}} \right) \, ,  \label{eq:evol2}
\end{equation}
where  we have introduced  a timescale for the magnetic torque,
\begin{equation}
\frac{1}{\tau_{\rm mag}}  = - \frac{1}{2 \tilde J_c}  \frac{d J_{mag}}{ dt} \,  .
\end{equation}

From equations~(\ref{eq:evol1}) and~(\ref{eq:evol2}) we can now derive a system of ordinary differential equations:
\begin{align}
& \frac{d \alpha}{dt} =  - \frac{2 \alpha}{\tau_{ \rm gw}}  - \frac{2 \alpha}{\tau_{ \rm v}} \frac{1+\alpha Q}{D} + \frac{2 P}{D} \frac{\alpha}{\tau_{\rm mag}}  \, , \label{eq:daldt} \\
& \frac{d \Omega}{dt} = \frac{2 F}{D} \left( \frac{\alpha}{\tau_{\rm v}} - \frac{1}{\tau_{\rm mag}} \right) \, ,  \label{eq:dOmdt}
\end{align}
where the total viscous damping time is defined as
\begin{equation}
\frac{1}{\tau_{\rm v} } = \frac{1}{\tau_{\rm s}} + \frac{1}{\tau_{\rm b}} \,  , 
\end{equation}
and the coefficients are given by the following expressions:
\begin{align}
& Q = \frac{d\tilde J_c}{d\Omega} \left( \frac{dJ_s}{d\Omega} \right) ^{-1} \, , \qquad  F = \tilde J_c \left( \frac{dJ_s}{d\Omega} \right) ^{-1} \, , \\
& P = \frac{d \ln \tilde E}{d\Omega} \cdot F \, , \qquad \qquad D = 1 + \alpha \left( Q - P \right) \, .
\end{align}
The form of equations~(\ref{eq:daldt}) and~(\ref{eq:dOmdt}) agrees with the evolution expected during the initial phase of the instability, 
in which the $f$-mode grows exponentially as $\alpha \sim e^{t/|\tau_{\rm gw}|}$ and the star slows down on viscous timescales.

As we have already pointed out, we expect that during the non-linear saturation regime 
the amplitude of the mode remains nearly constant~\cite{2004ApJ...617..490O, 2004PhRvD..70h4022S}. 
We can therefore approximate this evolution phase by assuming that  $d \alpha / dt = 0$, which allows us to recast equations~(\ref{eq:daldt})-(\ref{eq:dOmdt}) into a single relation, 
\begin{equation}
\frac{d \Omega}{dt} = - \frac{2  F}{ 1+\alpha Q } \left(  \frac{\alpha }{\tau_{\rm gw}} + \frac{1}{\tau_{\rm mag}} \right) \, . \label{eq:dOdt_sat}
\end{equation}
 This expression  shows that for neutron stars with a weak magnetic field, e.g. $B_{p} \leq 10^{11}$G (see Sec.~\ref{sec:MT}), the spin down  during the 
non-linear saturation phase is dominated, as expected,  by gravitational radiation.

Equations~(\ref{eq:daldt})-(\ref{eq:dOmdt}) and~(\ref{eq:dOdt_sat}) describe the time evolution of the mode amplitude and 
the star's angular velocity, but do not provide any information about the thermal evolution. 
In the next section, we address the cooling problem and 
 derive the equations for the thermal balance of the star.

\begin{figure*}
\begin{center}
\includegraphics[height=74mm]{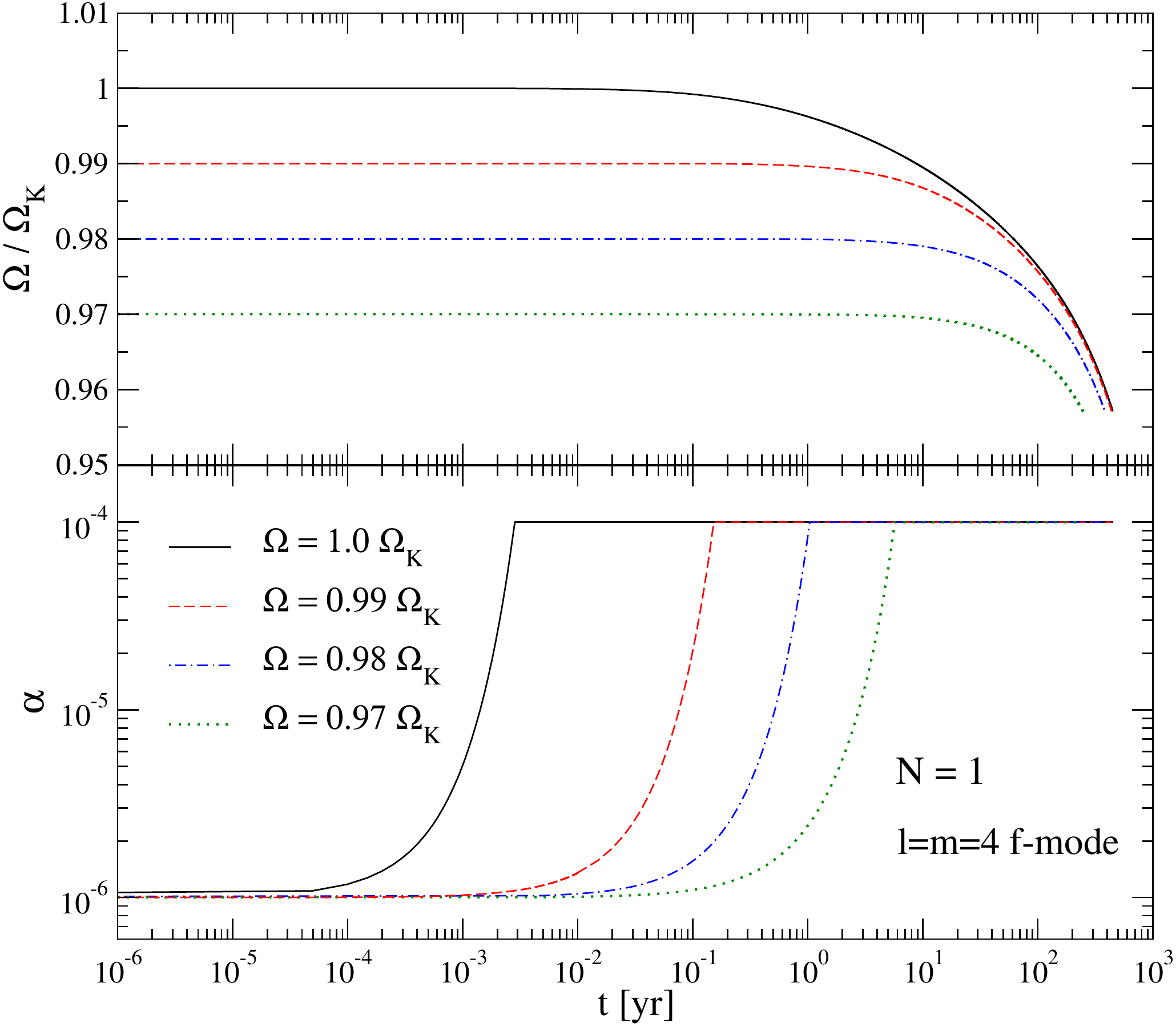}
\includegraphics[height=74mm]{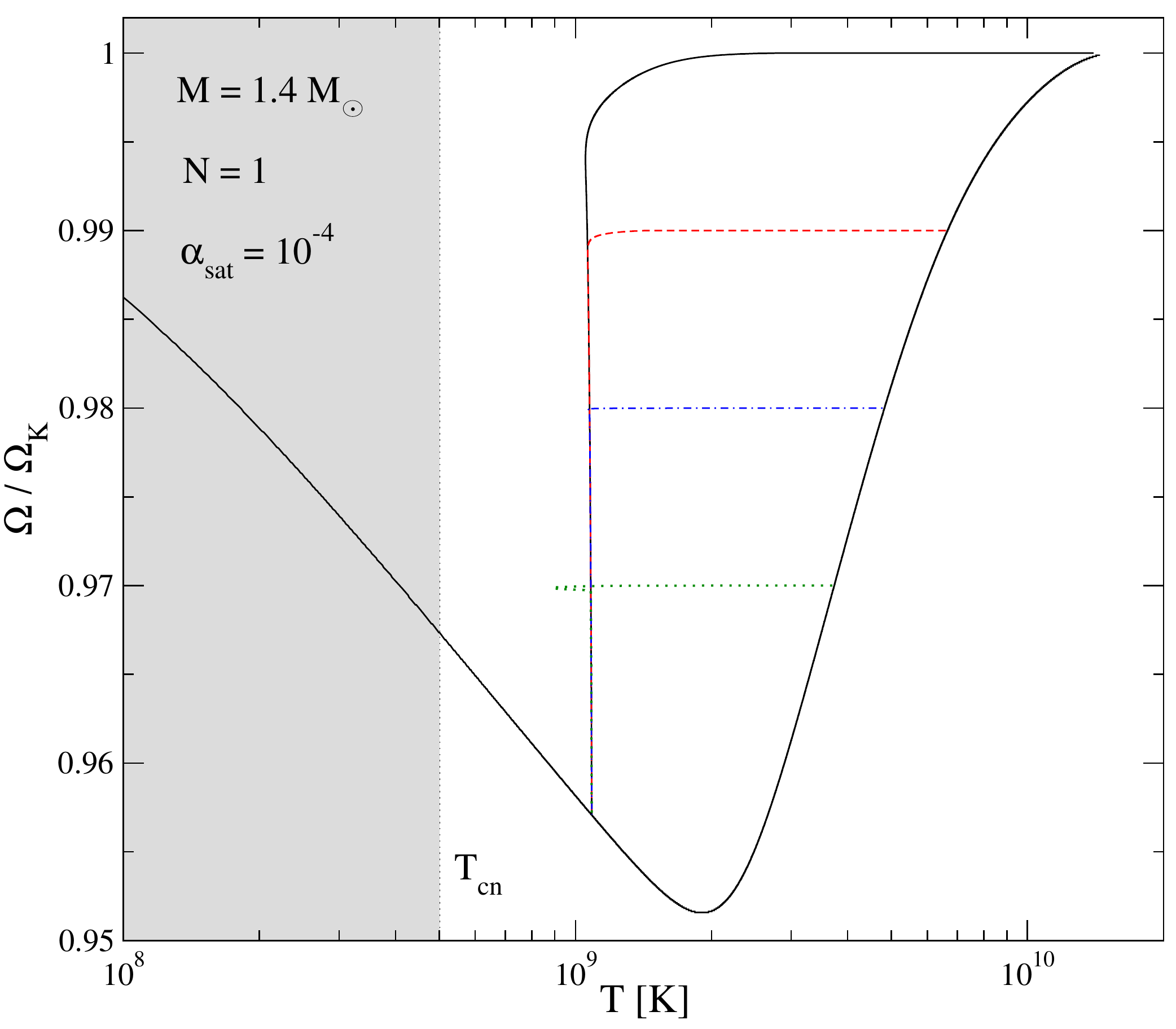}
\caption{Evolution of the $l=m=4$ $f$-mode instability for a relativistic polytrope with $N=1$, gravitational mass  $M =1.4\,M_{\odot}$ and $B_{\rm p } =10^{11}$~G. 
The star enters the instability window at different spins {\it (see right panel)} and correspondingly the left panel depicts the time-evolution of the  
stellar rotation rate {\it (top left panel)} and mode amplitude {\it (bottom left panel)}, where the initial amplitude $\alpha=10^{-6}$ saturates at $\alpha_{sat}=10^{-4}$. The shaded region in the right panel represents the temperature range where the 
neutrons and protons of the core are expected to be in a superfluid/superconducting state in accordance with recent models for the observed cooling of Cassiopeia A~\cite{2011PhRvL.106h1101P, 2011MNRAS.412L.108S}. 
\label{fig1}}
\end{center}
\end{figure*}

\subsection{Thermal evolution}

Few minutes after a core collapse, a neutron star becomes transparent to neutrinos 
which dominate the cooling for at least 1000 years~\cite{1983bhwd.book.....S}. 
However, an isolated neutron star may be re-heated by viscosity and magnetic field decay in later stages of its life. 
The evolution of the magnetic field  in the core is not well known in stars younger than $10^4$~yr~\cite{2012MNRAS.422.2632H}, and it is 
typically described by approximate relations~\cite{2009MNRAS.398.1869D,2012MNRAS.422.2878D,2012MNRAS.422.2632H}. 
For instance, an equation used in~\cite{2012MNRAS.422.2632H} reads 
\begin{equation}
B = B_0 \left(  1+ \frac{t}{\tau_{dec}} \right)^{-1} \, ,
\end{equation} 
where $B_0$ is the initial magnetic field and $\tau_{dec}$ the field decay timescale. 
This  is approximately $\tau_{dec} \approx 10^{4}$~yr and it is much longer than the typical evolution of the $f$-mode instability.  
Hence, the magnetic field decay can be reasonably neglected during the unstable phase of the $f$-mode.

The two viscous processes that operate in our models have an opposite effect on the thermal evolution. 
Bulk viscosity dissipates the mode energy by neutrino emission, as a result of the Urca reactions that 
attempt to restore the beta equilibrium in a displaced fluid element. These neutrinos induced by 
mode oscillations escape at infinity and cool down the star. However, this effect is very small compared 
to the dominant cooling rate of the background star and can be therefore neglected. 
More important is the shear viscosity which dissipates energy through particle collisions.  
The heat generated by shear viscosity is proportional to the mode amplitude and it can be 
 relevant during a mode instability. In fact as we will show in Section~\ref{sec:res}, shear viscosity may reduce and even completely balance 
the neutrino cooling of an unstable star. 

\begin{figure*}
\begin{center}
\includegraphics[height=75mm]{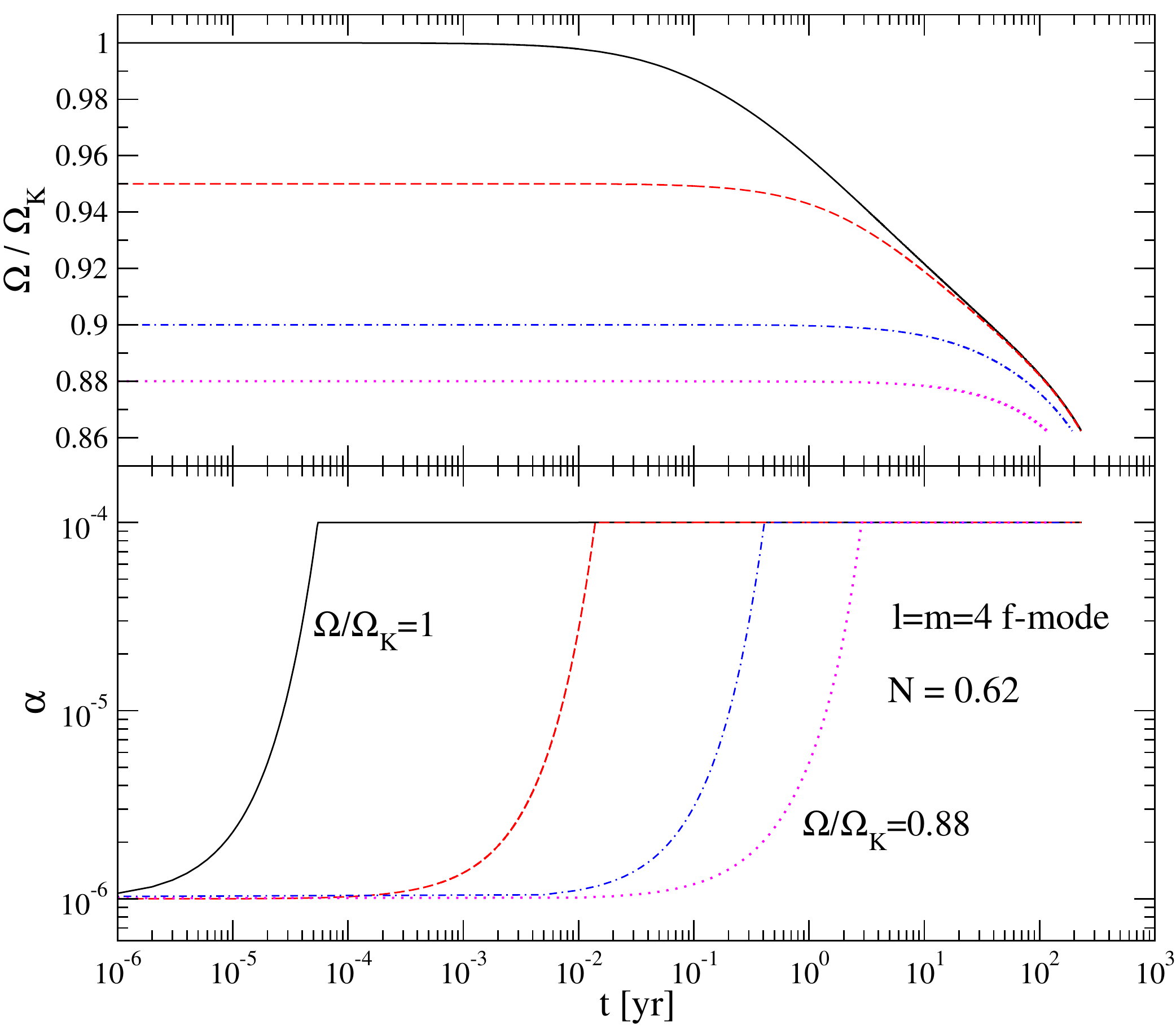}
\includegraphics[height=75mm]{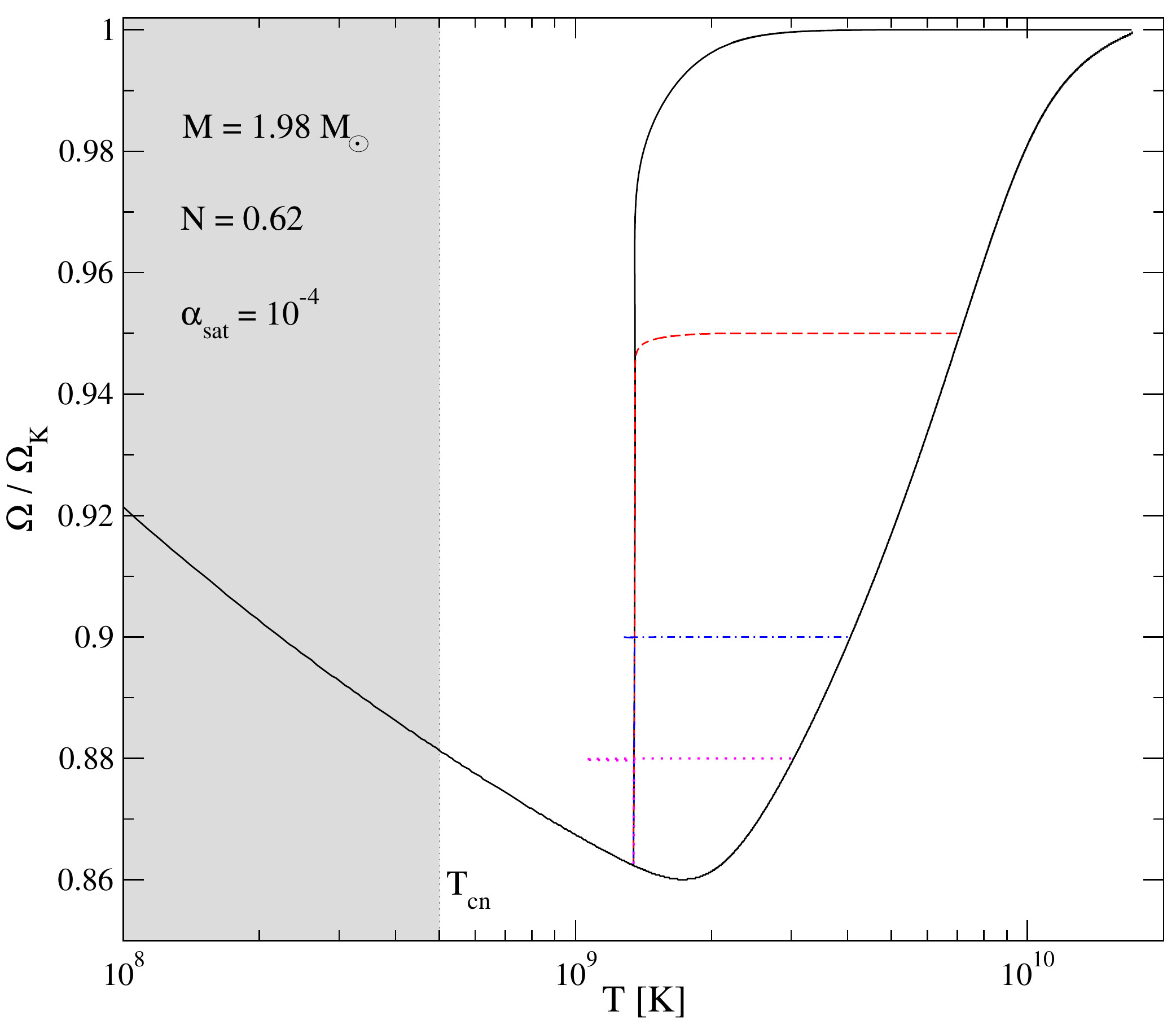} 
\caption{Evolution of the $l=m=4$ $f$-mode instability for a relativistic polytrope with $N=0.62$, mass $M =1.98 \,M_{\odot}$ and $B_{\rm p } =10^{11}$~G. 
The various panels depict the same quantities as described 
in Fig.~\ref{fig1}.
\label{fig2}}
\end{center}
\end{figure*}

The thermal evolution of a neutron star can be studied with a global  
energy balance between the relevant radiative and viscous processes~\cite{1983bhwd.book.....S}:
 \begin{equation}
C_v \frac{dT}{dt} = - L_{\nu} + H_s \, ,  \label{eq:Tevol}
\end{equation}
where $C_v$ is the total heat capacity at constant volume, $L_{\nu}$ is the neutrino's 
luminosity produced by the modified Urca processes that operate in the background star, and $H_s$ is the heating rate generated 
by shear viscosity. Although equation~(\ref{eq:Tevol}) is strictly valid  for isothermal neutron stars we consider it 
as a reasonable approximation for this work. 

The heat capacity depends in general on the neutron star EoS and temperature. 
An expression frequently 
used in literature is given by~\cite{1983bhwd.book.....S}   
\begin{equation}
C_v = 1.2 \times 10^{39} \frac{M}{M_{\odot}} \left(  \frac{\rho}{\rho_0} \right)^{-2/3} \frac{T}{10^9 \rm{K}}~\rm{erg}~\rm{K}^{-1} \, ,
\end{equation}
which has been strictly derived  for an ideal Fermi gas of neutrons in which 
  $\rho \lesssim 2 \rho_0$, where $\rho_0$ is the nuclear saturation density. 

The neutrinos' luminosity can be determined by the following equation~\cite{1983bhwd.book.....S}: 
\begin{equation}
L_{\nu} = 5.3 \times 10^{39} \frac{M}{M_{\odot}} \left(  \frac{\rho}{\rho_0} \right)^{-1/3} \left( \frac{T}{10^9 \rm{K}} \right)^{8} ~\rm{erg}~\rm{s}^{-1} \, ,
\end{equation}
which has already been used for the $r$-mode instability~\cite{1998PhRvD..58h4020O}.

The last term we need to specify in equation~(\ref{eq:Tevol}) is the heating rate induced by shear viscosity,   which 
 can be described by the following equation~\cite{1998PhRvD..58h4020O}: 
\begin{equation}
H_s = \frac{2 E}{\tau_{\rm s}} \, . \label{eq:Hs}
\end{equation}

We now have all the ingredients to study the evolution of the $f$-mode instability with 
 the system of equations~(\ref{eq:daldt})-(\ref{eq:dOmdt})  and~(\ref{eq:Tevol}).

\section{Results} \label{sec:res}

In this Section, we discuss the dynamical evolution of the $f$-mode  with and without the impact of an additional magnetic field. 
Furthermore, we study also the effect of an unstable $r$-mode on the evolution of the $f$-mode instability. 
For this, we model the compact objects as relativistic neutron stars which obey a polytropic EoS, 
\begin{equation}
p = K \rho^{1+1/N}
\end{equation}
where $p$ is the fluid pressure,
 $K$ is the polytropic constant and $N$ is the polytropic index. The rest-mass density $\rho$ is related to the fluid energy density $\varepsilon$ 
via the relation $\varepsilon = \rho + N p $. 

To study the dependence of the $f$-mode instability on the neutron star model, we consider  
more in detail two sequences of uniformly rotating stars up to the mass shedding limit $\Omega_{ \rm K}$. 
 The first model represents a  neutron star with polytropic index $N=1$.  
The  non-rotating member of this sequence has a gravitational mass of $M =1.4\,M_{\odot}$ and circumferential radius of $R=14.15~\textrm{km}$, while 
 the maximum rotating configuration has $\nu_{\rm K} = \Omega_{ \rm K} / 2 \pi = 673.1~\textrm{Hz}$  and $T/|W|=0.095$.  
The second model is described by a polytrope with $N=0.62$, which has a non-rotating configuration with 
$M = 1.98\,M_{\odot}$ and $R=11.95~\textrm{km}$. This star at the mass shedding limit rotates 
with $\nu_{\rm K} = \Omega_{ \rm K} / 2 \pi = 1086~\textrm{Hz}$ and has $T/|W| = 0.139$. 

 In this work we consider for simplicity only uniformly rotating stars. This is  however a reasonable assumption, as 
it is expected that magnetohydrodynamical processes  should redistribute the angular momentum inside a  proto-neutron star on few tens of rotation periods~\cite{Duez:2006qe}.

\begin{figure*}
\begin{center}
\includegraphics[height=75mm]{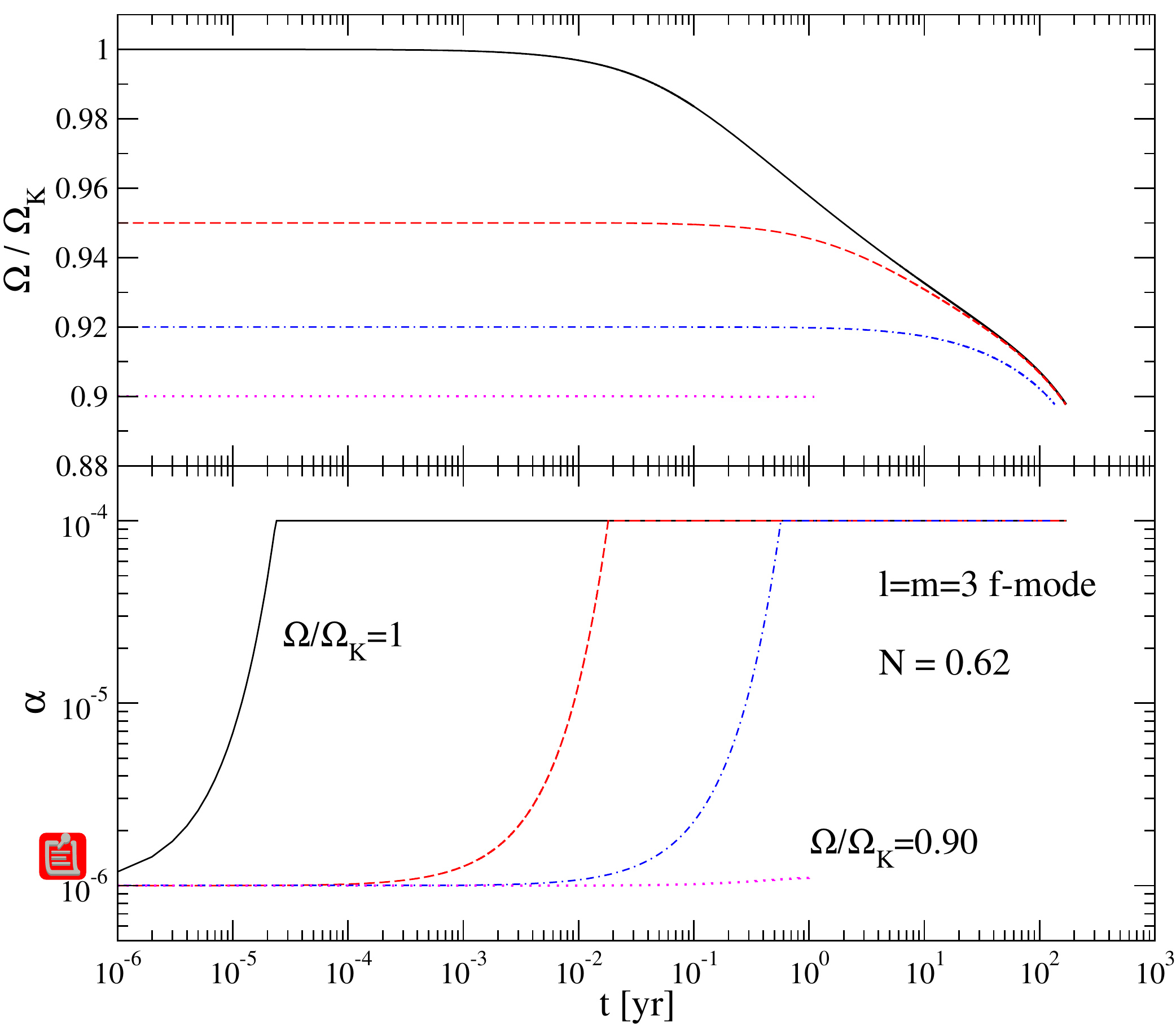} 
\includegraphics[height=75mm]{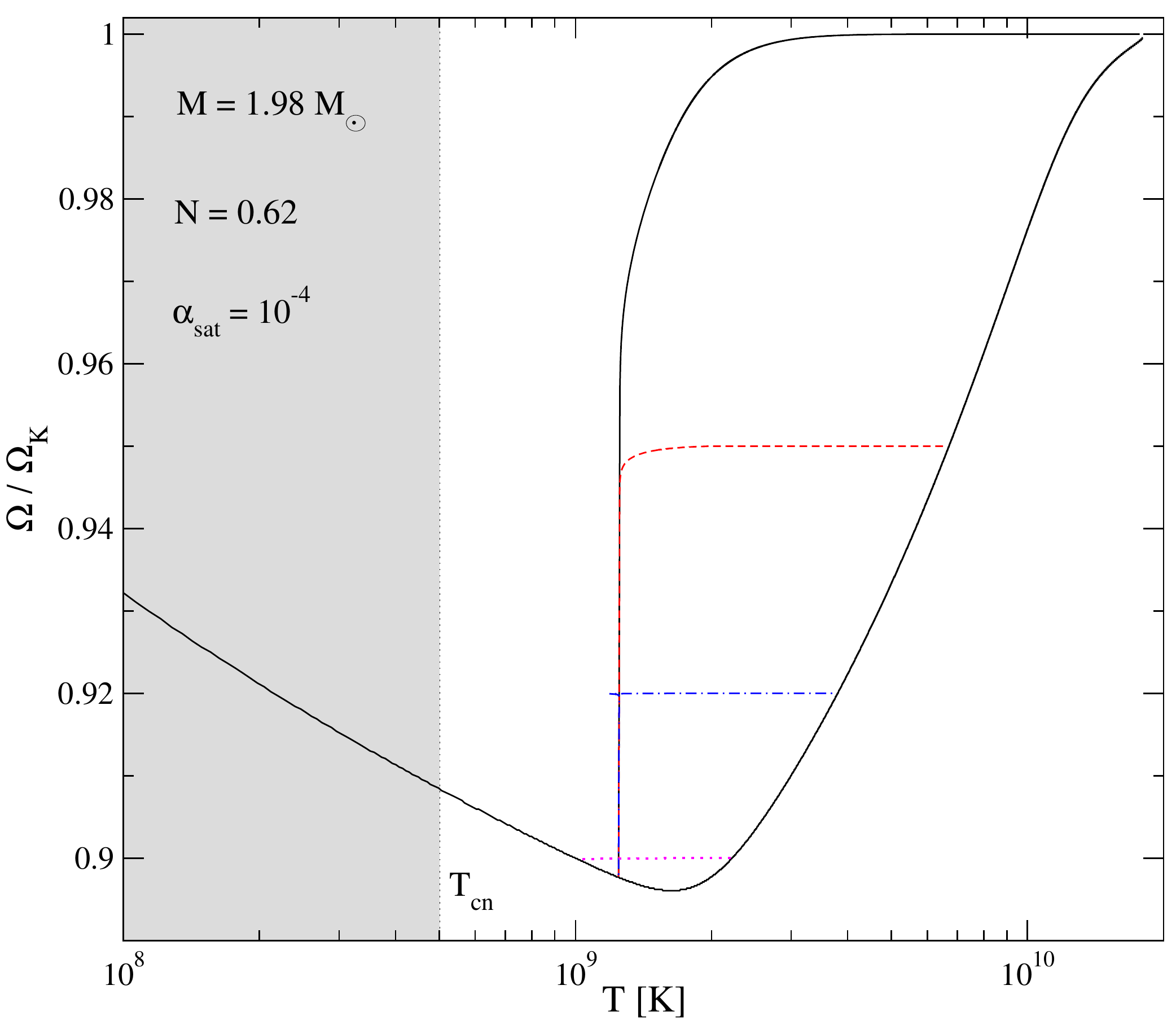}
\caption{Evolution of the $l=m=3$ $f$-mode instability for a relativistic polytropic star with $N=0.62$, 
 mass of $M =1.98\,M_{\odot}$ and $B_{\rm p } =10^{11}$~G. The various panels depict the same quantities as described 
in Figures~\ref{fig1} and~\ref{fig2}.
\label{fig3}}
\end{center}
\end{figure*}

In order to extract physical results from our simulations we must normalise the mode amplitude $\alpha$. 
In this work, we use the following definition
\begin{equation} 
E = \alpha  E_{\rm rot} \, ,  \label{req:alpn}
\end{equation}
where $E_{\rm rot}$ is the stellar rotational energy.  
For our polytropic models an $\alpha = 1$  corresponds to $E \sim  10^{-2}\,M_{\odot} c^2 $. 
The definition~(\ref{req:alpn}) implies that the fluid variables and the gravitational-wave strain scale as $\alpha^{1/2}$. 

The numerical scheme used to study the evolution of the $f$-mode instability is based on a 
fourth order Runge-Kutta algorithm. The numerical code evolves separately the two phases of the 
instability. After having specified  the initial conditions for  $\alpha, \Omega$ and $T$, we evolve 
 the initial growth phase of the $f$-mode according to equations~(\ref{eq:daldt})-(\ref{eq:dOmdt}).  
When the mode amplitude 
reaches a predetermined saturation value $\alpha_{sat}$,  the non-linear saturation phase is studied with relation~(\ref{eq:dOdt_sat}) 
until the star leaves the instability window and the numerical simulation ends. 
In both phases of the instability we use equation~(\ref{eq:Tevol}) 
for the thermal evolution.

\subsection{The evolution of the $f$-mode instability} \label{sec:FM}

We first consider the evolution of the $f$-mode instability in a low magnetised star ($B_p \leq 10^{11}$ G), 
and study the effect of the magnetic torque and $r$-mode in the following two Sections. 

The ideal conditions for the $f$-mode instability to work are high temperatures and rotation rates.  
At birth, the temperature of a  neutron star is around $T\simeq 10^{11}~\textrm{K}$  and drops down rapidly in the following few days. 
We therefore expect that as a rapidly rotating star cools down, it enters the $f$-mode instability window 
from the high temperature side (see Figures~\ref{fig1} and~\ref{fig2}). 
In our numerical scheme, we prescribe this behaviour by choosing appropriate initial conditions for $\Omega$ and $T$. 
In particular, for a given rotation rate $\Omega$ we specify the highest temperature allowed by the $f$-mode instability (e.g. see again Figures~\ref{fig1} and~\ref{fig2}).
The initial mode amplitude is set to $\alpha = 10^{-6}$; this
corresponds to an energy of $E\sim 10^{-8}\,M_{\odot} c^2$ which is a
typical value for the total 
energy loss in gravitational waves due to quadrupole deformations
shortly after a gravitational 
core collapse (see for instance~\cite{2008PhRvD..78f4056D, 2010A&A...514A..51S, 2012arXiv1210.6984M}).  
Although the initial pulsation energy of the 
$l=m=3$ and 4 $f$-modes might  be smaller, our results do not change 
significantly for different initial values of $\alpha$, as during the initial phase of the instability the mode's amplitude 
grows exponentially.

Another parameter one needs to specify in the evolution scheme is the saturation amplitude of the $f$-mode. 
From non-linear dynamical simulations one can determine  upper limits on the 
quadrupole $f$-mode~\cite{2010PhRvD..82j4036K}, but the problem is still open for higher multipole $f$-modes. 
We then consider the saturation amplitude as a free parameter. In our simulations  we choose $\alpha_{sat} = 10^{-4}$, which 
is equivalent to $E\sim 10^{-6}\,M_{\odot} c^2$. 
These values are ``reasonable''  and much smaller than what is frequently 
used in literature~\cite{1995ApJ...442..259L, 1998PhRvD..58h4020O}, where $\alpha_{sat}=1$.

For the $N=1$ model we consider four cases for the $l=m=4$ $f$-mode instability  with different initial rotation rates and temperatures. 
Figure~\ref{fig1} shows the results for the star's spin, the mode's amplitude and the star's trajectory through the instability window. 
The entire evolution lasts for about 500 years, while the duration of the initial growth phase depends on the 
initial rotation rate. For a star that becomes unstable at $\Omega = \Omega_{ \rm K}$ ($\Omega = 0.97\,\Omega_{ \rm K}$) the growth phase lasts for about 1 day (6 years).
This phase is dominated by the growth time $\tau_{\rm gw}$ which increases gradually for slower rotating stars, from $\tau_{\rm gw} \simeq 10^{4}~\textrm{s}$ 
at the mass shedding limit to  $\tau_{\rm gw} \simeq 10^{8}-10^{9}~\textrm{s}$ at the bottom of the instability window, $\Omega \sim 0.95\,\Omega_{ \rm K} $~\cite{2011PhRvL.107j1102G}. 
After the mode saturates, Figure~\ref{fig1} (right panel) shows that 
 the star spins down at nearly constant temperature ($T \simeq 10^{9}$ K) as a result of the heat generated by shear viscosity which completely balances  
 the neutrino cooling.  
\emph{This  effect prevents the star from entering the superfluid transition zone and increases the duration of the instability.} 
In fact, for $T \leq T_{\crt\n}$ the neutrons of the core become superfluid and mutual friction damps the  
$f$-mode very efficiently on short timescales~\cite{1995ApJ...444..804L}. 
For the superfluid critical temperature we chose a value of $T_{\crt\n}\simeq 5 \times 10^{8}\ \textrm{K}$ which has recently been determined from the cooling curves  
of Cassiopeia~\cite{2011PhRvL.106h1101P, 2011MNRAS.412L.108S}.

In Figure~\ref{fig1} (right panel) it is noticeable that the trajectory of a star which becomes unstable at $\Omega = 0.97\,\Omega_{ \rm K}$ is slightly different from the other cases. 
This is due to the slower growth phase 
of the mode which delays the effect of the shear viscosity heating on the thermal evolution. Therefore, 
the star first cools down to a minimum temperature, 
then is re-heated by shear viscosity and finally evolves quasi-isothermally in the last  part of the trajectory. 

The evolution of the $l=m=4$ $f$-mode instability for the $N=0.62$ model  is qualitatively similar to the $N=1$ case, but it 
 develops faster  and within a significantly larger instability window (see right panel of Fig.~\ref{fig2}).   
At the mass shedding limit the  gravitational growth time is  only  $\tau_{\rm gw} \simeq  700 \, \text{s}$ in contrast to the 
$\tau_{\rm gw} \simeq  10^4 \, \text{s}$ for the $N=1$ model,  and 
the total evolution now lasts for about 200 years. The exponential growth of the $f$-mode takes 
about 26 minutes (5 days) for a star with an initial $\Omega = \Omega_{ \rm K} $ ($\Omega = 0.95\,\Omega_{ \rm K} $).

For the $N=0.62$ model we have a relevant instability window also for the  $l=m=3$ $f$-mode. In fact, at the Kepler limit  
the gravitational growth time, $\tau_{\rm gw} \simeq  300~\text{s}$, is even shorter than the $l=m=4$ $f$-mode 
but increases more rapidly with decreasing rotation. 
The resulting instability region is shown in Figure~\ref{fig3} for  a star that becomes unstable at different rotation rates. 
The total instability evolution time is about 200 years and the duration of the  growth phase is comparable to the $l=m=4$ $f$-mode case. 
Furthermore, Figure~\ref{fig3} shows that when a star becomes unstable at lower rotation rates near the minimum of the instability window, 
 e.g. $\Omega \simeq 0.90\,\Omega_{ \rm K} $, the mode amplitude remains small and the shear viscosity heating does not balance the neutron star's cooling.

The properties of the instability window and gravitational growth time make the $N=0.62$ massive star  a better gravitational-wave 
source than the $N=1$ model (see Section~\ref{sec:GW}).

\subsection{Magnetic torque} \label{sec:MT}

The main mechanism that gradually slows down a neutron star is electromagnetic radiation, which 
is powered by the star's rotational energy. Since the magnetic torque increases considerably with rotation, 
as shown by the dipole formula~(\ref{eq:dJmagdt}), the magnetic braking  may affect the evolution of the $f$-mode instability 
by accelerating the spin-down.
This means that  an unstable $f$-mode may not have the time to grow significantly and generate a detectable gravitational-wave signal.

We study the impact of the magnetic field on the $l=m=4$ $f$-mode instability for the  $N=1$ and $N=0.62$  models.  
For each model we consider two dipolar magnetic field configurations with, respectively,  $B_{\rm p }=10^{12}$G and $B_{\rm p }=10^{13}$G, 
and  focus on an orthogonal rotator (see Section \ref{sec:modevol}). 

\begin{figure}[!h]
\begin{center}
\includegraphics[height=76mm]{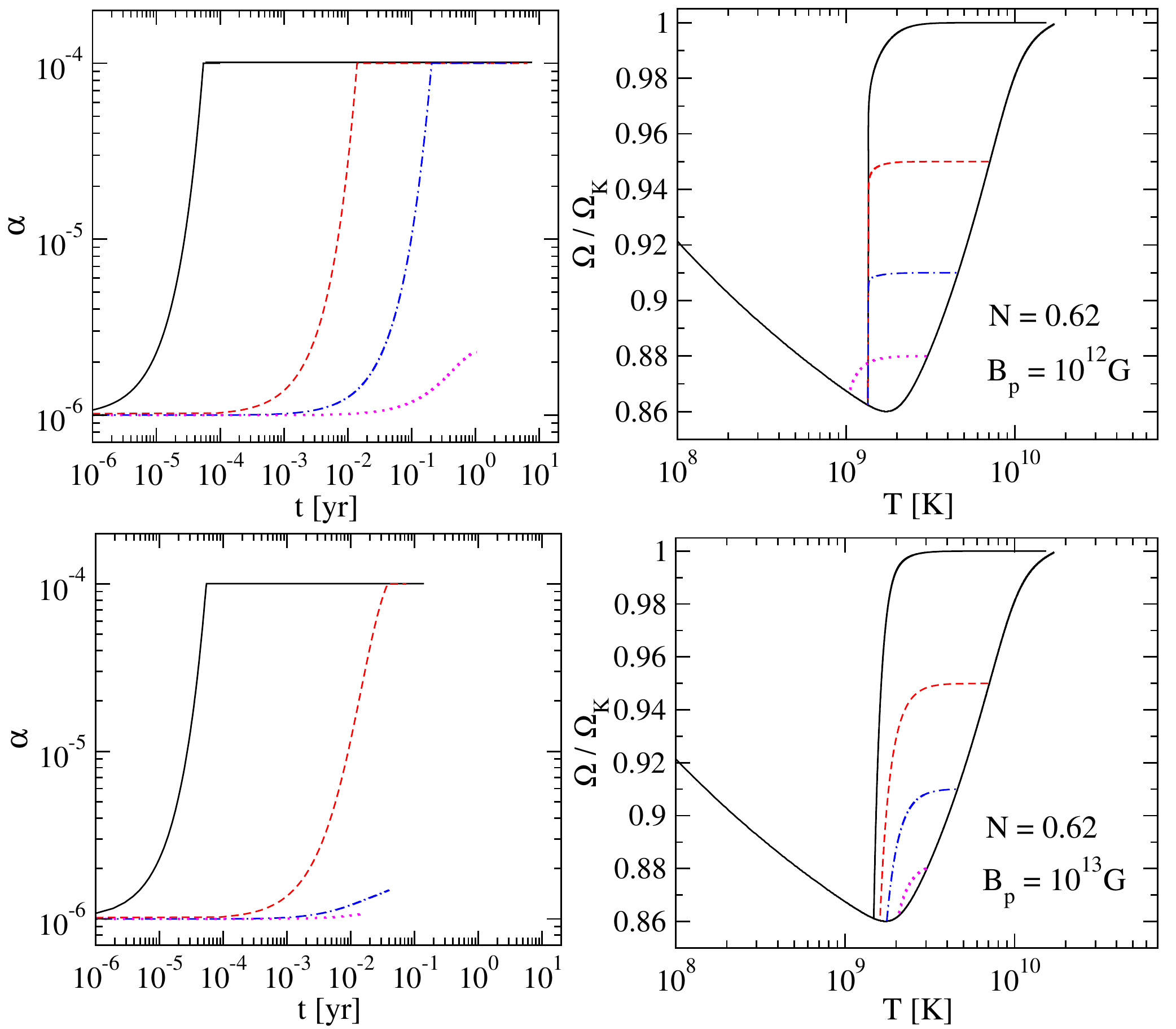} 
\caption{ 
The impact of the magnetic torque on the evolution of the $l=m=4$ $f$-mode instability for the $N=0.62$ model. 
The magnetic field is dipolar and orthogonal to the rotation axis. 
The two top panels display the mode amplitude (left panel) and 
the star's evolution through the instability region (right panel) for $B_{\rm p }=10^{12}$ G. 
The same quantities are depicted in the two lower panels for $B_{\rm p }=10^{13}$ G. 
\label{fig4}}
\end{center}
\end{figure}

We evolve various initial conditions with different 
temperatures and rotation rates.  The evolution of the mode amplitude and the star spin-down is shown 
in Figure~\ref{fig4} for the $N=0.62$ model. 
For a magnetic field of $B_{\rm p }=10^{12}$ G the total duration of the instability is shorter than a factor of about $20$ with respect to the 
 $B_{\rm p }=10^{11}$G model.
In fact, the evolution lasts about $10$~yr for the $N=0.62$ and $N=1$ models. 
In slower rotating models, the magnetic torque clearly dominates the evolution 
even during the initial phase of the instability and limits considerably the growth of the mode amplitude. 
This behaviour is for instance evident in Figure~\ref{fig4} for a $N=0.62$ model with an initial rotation rate of $\Omega = 0.88\,\Omega_{ \rm K}$.  
The magnetic torque dominates completely the evolution when $B_{\rm p }=10^{13}$ G. The unstable star is quickly spun down 
and the mode amplitude is strongly limited even in the more massive model (see Figure~\ref{fig4}).

\subsection{The $f$-mode versus the $r$-mode instability} \label{sec:RM}

\begin{figure*}[t]
\begin{center}
\includegraphics[height=75mm]{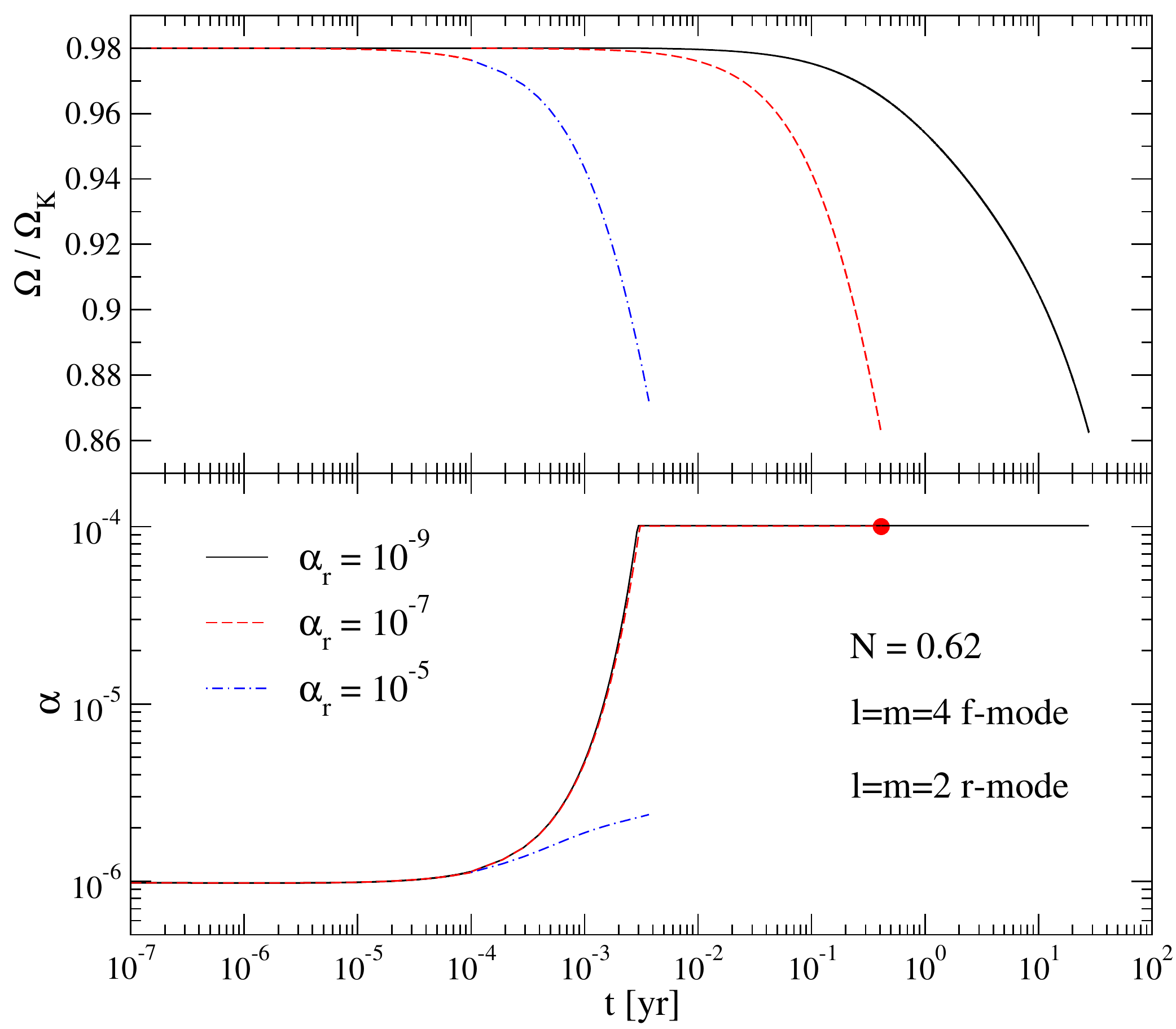} 
\includegraphics[height=75mm]{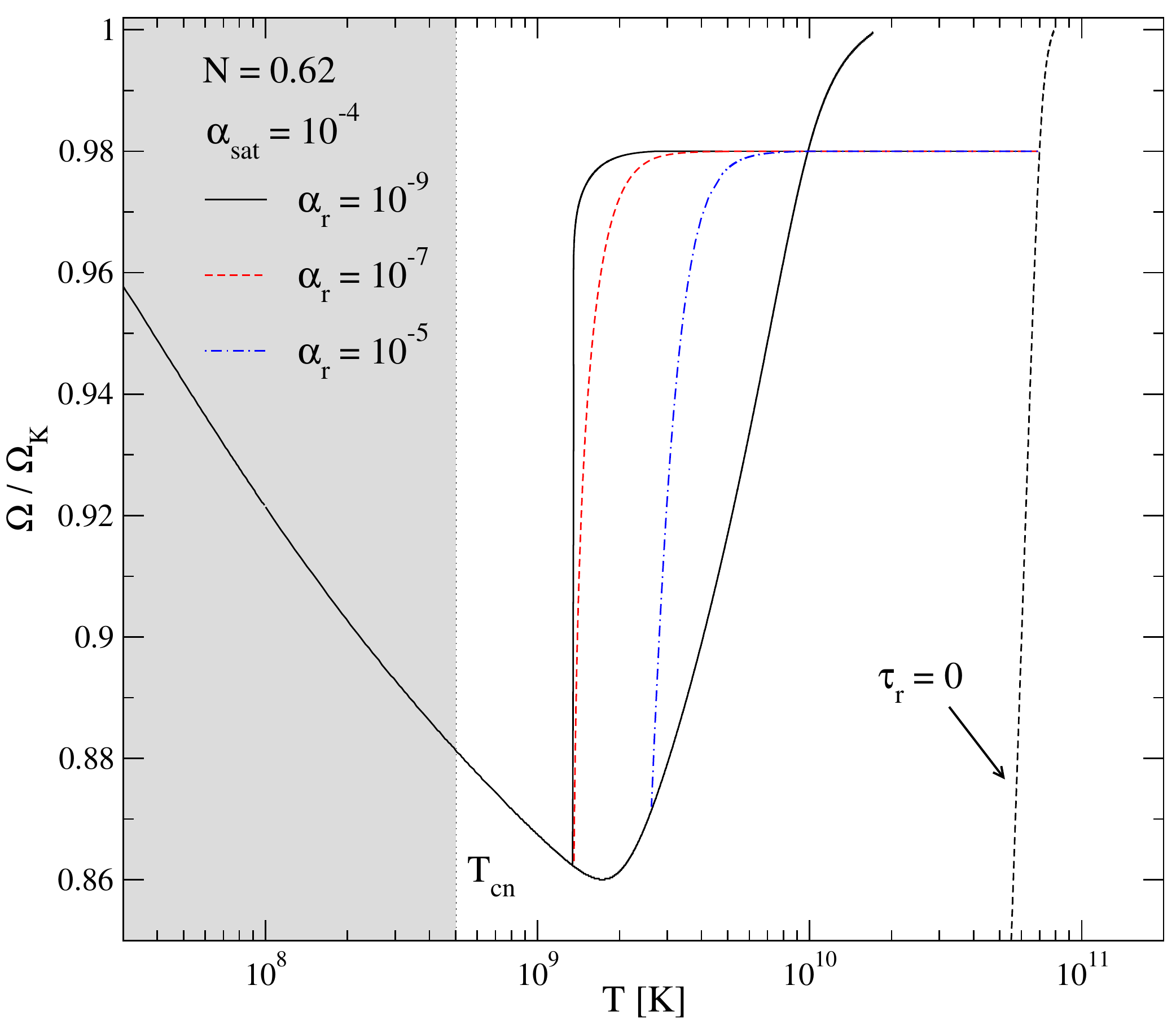} 
\caption{The effects of an unstable $l=m=2$ $r$-mode on the evolution of the $l=m=4$ $f$-mode instability for a relativistic polytrope with 
$N=0.62$. The saturation amplitude of the r-mode is varied between $\alpha_{r} = 10^{-9}-10^{-5}$ 
while the maximum amplitude of the $f$-mode is kept constant at $\alpha_{sat} = 10^{-4}$. The left panel shows the evolution of the stellar rotation (top panel) and the $f$-mode amplitude 
(lower panel) for the various examples. The filled-circle  in the lower left panel denotes the end of the f-mode instability 
for $\alpha_{r} = 10^{-7}$. The right panel shows the evolution through the instability window, where 
the dashed line denotes the high temperature part of the $r$-mode critical curve ($\tau_{r}=0$).  
\label{fig6}}
\end{center}
\end{figure*}

In rapidly rotating neutron stars several modes can be driven unstable 
by gravitational radiation at the same time. The most important ones are definitely the 
$f$- and $r$-modes, as they have a comparatively short growth time and therefore can generate 
a significant gravitational wave signal.  A relevant difference between these two classes of modes is 
that the $f$-mode only gets unstable in very rapidly rotating stars while the $r$-mode is CFS unstable 
at any rotation rate. In addition, the growth time is typically  shorter for the $r$-mode which consequently has a larger instability window. 
It is then reasonable to think that the $r$-mode  
should dominate the evolution of the gravitational wave driven instability. 
However before drawing any secure conclusion, 
it is necessary to know  the maximum  amplitude that each mode can reach during the instability.

Non-linear perturbation calculations show that the $r$-mode may  transfer energy to other inertial modes 
through non-linear mode coupling. 
As a result the maximum amplitude of the $r$-mode may be limited to $c_{\al} = 10^{-3}-10^{-5}$,  
where $E= c_{\al}^{2}  E_{\rm rot}$~\cite{2002PhRvD..65b4001S,2007PhRvD..76f4019B}.   
In our notation, these values correspond to a saturation 
amplitude of  $\alpha_r =  c_{\al}^{2}=10^{-6} - 10^{-10}$, since we parametrize the mode energy according to $E=\alpha_r E_{\rm rot}$. 

As it was already pointed out, the maximum amplitude of the $f$-mode is still uncertain. This forces us to consider  
a ``reasonable'' value for the $f$-mode saturation which we set to
$\alpha_{sat} = 10^{-4}$; the same value that was used in the previous
sections. 
We then study the combined evolution of $f$- and $r$-mode for different $r$-mode saturation amplitudes.
In order to address this problem we derive an additional set of equations which  are given in the Appendix.

We focus on the $l=m=2$ $r$-mode and the $l=m=4$ $f$-mode,  which are the most relevant unstable modes 
in  their respective classes  due to their relatively short instability growth time with respect to the 
dissipative processes~\cite{2003CQGra..20R.105A}.  
We consider the more massive stellar model with $N=0.62$ as it is potentially a better source for gravitational-wave detection, and 
neglect for simplicity the effect of the $r$-mode on the magnetic field. As shown in~\cite{2000ApJ...531L.139R}, the toroidal magnetic field component 
may be amplified by an $r$-mode and this can change the evolution of the instability. In order to assess this effect it would be necessary to study also the 
back-reaction of the magnetic field on the mode and determine a maximum magnetic field amplification.

The properties of the $f$-mode have been already determined in Section~\ref{sec:FM}, while for the $r$-mode one needs to consider some approximations 
of its viscous damping and gravitational radiation growth times. More precisely, the various  dissipative timescales are determined via the
analytical  relations given in~\cite{1999A&A...341..110K} which have been derived for a uniform density star.  
For the model under consideration here, the $l=m=2$ $r$-mode growth time at the mass shedding limit is $\tau_{\rm gw} \simeq 2.9~\textrm{s} $, which is more than two orders 
of magnitude smaller than the $l=m=4$ $f$-mode growth time ($\tau_{\rm gw}  = 700~\textrm{s}$).  
The oscillation frequency  in the rotating frame of the $l=m=2$ $r$-mode is determined  at leading order by $\omega = 2 \Omega /3$, 
and for simplicity it is also assumed that the canonical angular momenta of the $r$- and $f$-mode are equal. 
These approximations prevent us from determining the properties of the $r$-mode from numerical evolutions. 
However,  since the $r$-mode grows much faster than the $f$-mode it is not expected that the qualitative behaviour of the results 
will change considerably with a more accurate description of the $r$-mode.

As the star cools down, it is expected that the $r$-mode gets unstable before the $f$-mode as its instability window extends towards higher temperatures  (see Figure~\ref{fig6}). 
The $r$-mode then  quickly reaches its non-liner saturation 
value $\alpha_{r}^{sat}$ and spins down the star.  
However  if  the $r$-mode growth is limited by non-linear mode coupling,  
the star evolves towards the critical curve of the $f$-mode and eventually also this mode is driven unstable by gravitational radiation.

In Figure~\ref{fig6} we show a representative case in which the star enters the critical curve of the $r$-mode at 
$\Omega = 0.98\,\Omega_{ \rm K}$ with an initial amplitude  $\alpha_r = 10^{-10}$. We consider several simulations 
where the $r$-mode grows up to a maximum amplitude of $\alpha_{r}^{sat}=10^{-9}-10^{-5}$, while for the $f$-mode  we set 
$\alpha_{sat} = 10^{-4}$. 
The impact of the $r$-mode on the star's spin-down and the total evolution of the $f$-mode is evident already at low saturation $\alpha_{r}$.  
The star leaves the instability window of the $f$-mode after an evolution time given by 
$t_{ev} \sim 300 \times  \left( 10^{-9} / \alpha_{r}^{sat} \right) \textrm{yr}$.  When $\alpha_{r} \geq 10^{-5}$ the evolution is completely 
dominated by the $r$-mode and the $f$-mode amplitude is strongly constrained to small values. 
However, for $\alpha_{r} \leq 10^{-7}$ the $f$-mode still has time to grow and potentially generates a detectable 
gravitational-wave signal (see Section~\ref{sec:GW}). As expected, it is therefore crucial to know more accurately the relative saturation amplitude 
between  these two modes. This is an interesting aspect that must be clarified in a future work.

\section{Gravitational Waves} \label{sec:GW}

The instability of the $f$-mode is potentially a strong source of gravitational radiation and it is then logical 
to estimate the detectability prospectives by the current and next generation of Earth-based laser interferometers. 

The two independent polarization states of the gravitational wave strain  can be expressed in terms of the 
spin-weighted spherical harmonic ${}_{-2}Y^{l m}$ as
\begin{equation}
h_{\theta \theta}^{l m} - i h_{\theta \phi}^{l m} = h^{l m} {}_{-2}Y^{l m} \, , \label{eq:hdef}
\end{equation}
where  $h^{l m}$ is defined as  
\begin{equation}
h^{l m} \equiv  \frac{G}{c^{l+2}}   \frac{\beta_{l} }{r}  \left( i \omega_I  \right)^{l}     \delta D_{l m}   \,     \label{eq:h44b}
\end{equation}
and  $\omega_I = \omega -m\Omega$ is the mode frequency in the inertial frame. The coefficient $\beta_l$ is given by 
\begin{equation}
\beta_l \equiv \frac{8 \pi }{l \left( l-1\right) \left( 2 l+ 1 \right) !! } \sqrt{ \frac{ \left(l+2 \right)! }{ \left(l-2 \right)!} } \, .
\end{equation}

A more suitable quantity to evaluate the gravitational-wave detection  is the characteristic strain, 
which also takes into account the 
statistical amplification of the signal due to the number of oscillations accumulated  in a given frequency bandwidth.
The characteristic strain is defined by the following equation:
 \begin{equation} 
h_{c}^{lm}  \equiv  \langle \left| h^{l m} {}_{-2}Y^{l m}  \right| \rangle \sqrt{ N_{cyc} }  \, , \label{eq:hc} 
\end{equation}
where  $\langle \dots \rangle$ denotes an averaging over the angles $\left( \theta, \phi \right)$.
  The number of oscillation cycles can be expressed as 
$N_{cyc} = \nu \, \tau_{ev} $, which is written in terms of  the mode frequency $\nu = \omega_I/2\pi$ and 
the time spent near a given frequency  
$\tau_{ev}$. 

\begin{figure*}[t]
\begin{center}
\includegraphics[height=75mm]{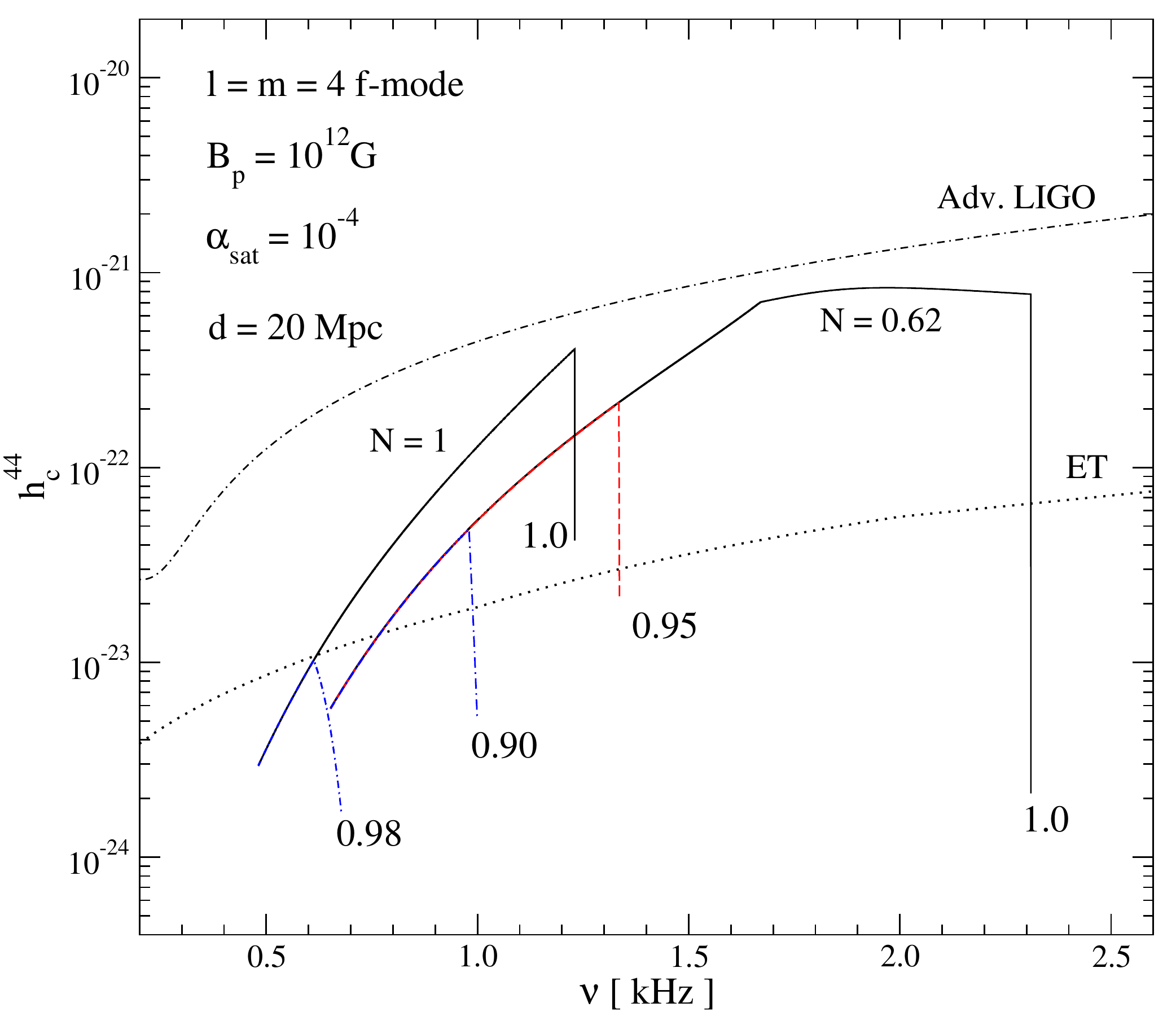}
\includegraphics[height=75mm]{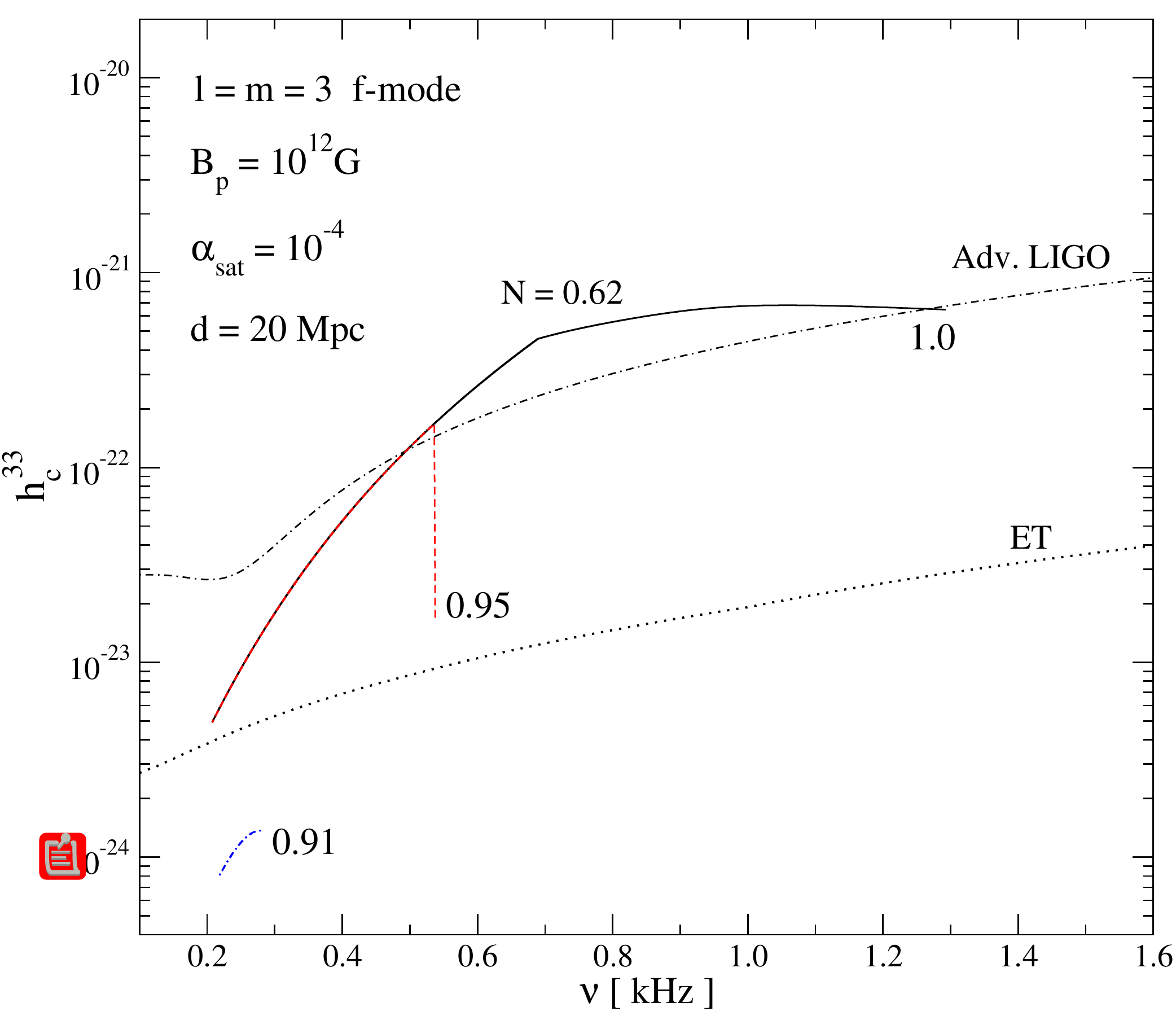} 
\caption{ 
The characteristic strain generated by the $f$-mode instability for a  
polytropic star with $B_{\rm p }=10^{12}$ G. The left panel shows the gravitational-wave signal of the $l=m=4$ $f$-mode for the 
$N=1$ and $N=0.62$ models, while the right panel depicts the signal of the $l=m=3$ $f$-mode for the  $N=0.62$ model.
The source is located at 20 Mpc and the saturation amplitude of the 
$f$-mode is set to $\alpha_{sat} = 10^{-4}$. The notation used in this Figure is the same as in Figure~\ref{fig0}. 
\label{fig7}}
\end{center}
\end{figure*}

For magnetic fields with a magnitude of $B_{\rm p} \le 10^{11}$ G, the impact of the magnetic torque on the $f$-mode instability is irrelevant~(see Section~\ref{sec:res}),  
and the mode evolves like in an unmagnetised star.
This means that during the initial exponential growth of the mode the 
star spins down on viscous timescales, i.e.\ $\Omega$ is nearly constant, and the radiated gravitational-wave signal is virtually monochromatic.  
As a result, the quantity $\tau_{ev}$ can be approximated by the accumulated evolution time of the star during this initial phase with $\Omega \simeq const$.  
After the $f$-mode saturates, the amplitude remains approximately constant and 
gravitational radiation removes angular momentum from the star.  
This leads to a frequency variation which can be well approximated by the following relation:
\begin{equation} 
\tau_{ev} =  \nu \left| \frac{ d t  }{ d \nu} \right|  \, . \label{tau_ev}
\end{equation}

When a star has higher magnetic field strengths, $B_{\rm p} > 10^{11}$ G,  the magnetic torque affects also the initial phase of the instability 
and accelerates the spin-down. The $f$-mode frequency therefore evolves as the  star loses angular momentum by magnetic 
braking and equation~(\ref{tau_ev}) has to be used throughout the entire instability evolution.  

However, realistic calculations of the characteristic strain also have to consider the maximum signal integration time allowed 
by the detector technology. 
We assume that a detector may integrate the signal at most for 1 year, and consequently    
calculate $h_{c}^{lm}$ for $\tau_{ev}  \leq 1 \textrm{yr}$.  This means that in $N_{cyc}$ we use  the evolution time of our simulations whenever $\tau_{ev} < 1 \textrm{yr}$, otherwise we set $\tau_{ev} = 1 \textrm{yr}$. \\

\begin{figure}
\begin{center}
\includegraphics[height=75mm]{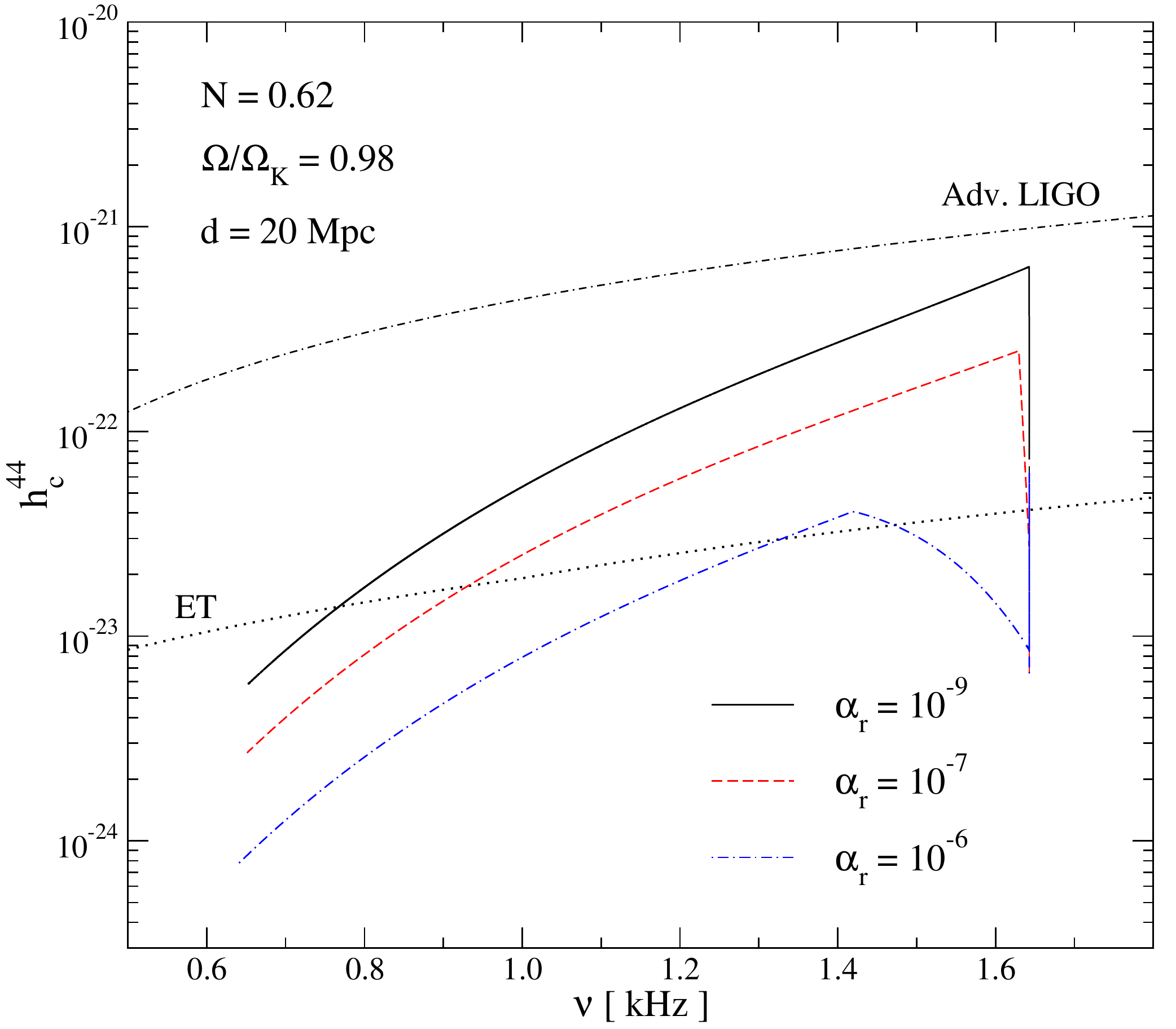} 
\caption{ 
The characteristic strain generated by the $l=m=4$ $f$-mode
instability when an $l=m=2$ $r$-mode is excited as well. 
The star is a relativistic polytrope with $N=0.62$, $B_{\rm p }=10^{11}$ G and is located at 20 Mpc. It enters the  $r$-mode 
instability window  at $\Omega = 0.98 \Omega_{\rm K}$. The Figure shows three cases with different 
$r$-mode   saturation amplitudes while the saturation amplitude of the 
$f$-mode is fixed to $\alpha_{sat} = 10^{-4}$.  
\label{fig8}}
\end{center}
\end{figure}

The results  for the $N=1$ and $N=0.62$ models are shown in Figure~\ref{fig0} and in Figures~\ref{fig7}, \ref{fig8} 
together with the sensitivity curves of Advanced LIGO and ET~\cite{LIGO, ETel}. 
The different noise curves of the detectors are determined by using $h_{\rm rms} = \sqrt{ \nu S_{h} (\nu)}$, where 
$S_{h} (\nu)$ is the power spectrum of the detector at hand.

For each model we determine the 
characteristic strain of the various evolutionary paths studied in Sec.~\ref{sec:res} in which 
the onset of the $f$-mode instability happens at different rotation rates. 
As shown in Figure~\ref{fig0}, the signal is initially monochromatic during the growth phase of the mode  
and evolves successively to lower frequencies as the star spins down.
We find that if  an unstable $f$-mode with multipoles $l=m=3, 4$ saturates at $\alpha_{sat}=10^{-4}$, 
the gravitational-wave signal can be detected by ET for the most part of the instability evolution for the more massive stellar model ($N=0.62$) located in the Virgo Cluster at a distance of 20~Mpc. 
 Actually, the $l=m=3$ $f$-mode may be detectable also by Advanced LIGO/Virgo. 
For the $N=1$ model 
an unstable $l=m=4$ $f$-mode may generate a detectable signal only if 
the star rotates  near the Kepler limit  $\Omega \gtrsim 0.98\, \Omega_{\rm K}$.

A higher magnetic field may affect significantly the characteristic strain of the $f$-mode, as the 
number of accumulated cycles $N_{cyc}$ decreases.  
The results for the $B_{\rm p} =10^{12}$ G case are shown in Figure~\ref{fig7} 
for both the $N=1$ and $N=0.62$ models.  Although the gravitational-wave signal is slightly weaker than in the previous $B_{\rm p} =10^{11}$ G case, it is 
still detectable by ET  and the $l=m=3$ $f$-mode of the $N=0.62$ model is also above the sensitivity curve of  Advanced LIGO.
The effect of the magnetic breaking is particularly evident if the star gets unstable at lower rotation rates. 
For instance, the characteristic strain of the $N=1$ model with an initial rotation rate of $\Omega = 0.97\, \Omega_{\rm K}$ is significantly smaller 
than for the low magnetised model. 
We  have also calculated the characteristic strain for models with a magnetic field strength of $B_{\rm p} =10^{13}$ G and found 
that the gravitational-wave signal of the $N=1$ model drops below the detectors' sensitivity curves. 
This happens also to the $N=0.62$ model when $\Omega \lesssim 0.96 \Omega_{\rm K}$. 
The results presented in this Section can be easily rescaled with the $f$-mode saturation amplitude as $h_{c} \sim \alpha_{sat}^{1/2}$.

If in addition a $r$-mode is present as well, the results for the gravitational wave strain depend on the value of the saturation amplitude for this auxiliary mode; see Figure~\ref{fig8} for the $N = 0.62$ polytrope. Supposing a small $r$-mode saturation, the expected signal can still be detected with ET from a source in the  Virgo cluster. However, as the saturation amplitude of the $r$-mode is increased, its instability dominates the evolution of the $f$-mode instability. The star loses angular momentum too fast for the $f$-mode to grow substantially before it leaves its instability window (see also Figure~\ref{fig6}), leading to a significant reduction in the gravitational wave strain for large $r$-mode saturation values. In these cases the signal of the $f$-mode will drop even below the ET sensitivity curve.
However, estimating the gravitational-wave growth time of the $r$-mode with a uniform density stellar model~\cite{1999A&A...341..110K}, we find that an unstable $l=m=2$ $r$-mode can  
 generate a gravitational-wave signal detectable by ET. 
This means that  when both the $f$- and $r$-modes are simultaneously excited and the $r$-mode saturation amplitude is $\alpha_{r}^{sat} \lesssim 10^{-7}$, 
 ET should be able to detect the gravitational-wave signal emitted by both these two modes. Instead, for larger $\alpha_{r}^{sat}$ the $r$-mode will be 
 still detectable but the $f$-mode signal should be below the ET sensitivity curve.

\section{Conclusions} \label{sec:conc}

We have presented in this work the first dynamical study of the $f$-mode instability which  considers relativistic rapidly rotating stars 
and incorporates the effect of viscosity, magnetic fields and  unstable $r$-modes. 
These are the most dominant effects which may have a 
significant impact on the evolution of this mode. 
 Our neutron star's models are relativistic and described by a polytropic equation of state. 
We consider more in detail two sequences of uniformly rotating stars which can rotate up to the Kepler limit and describe, 
 respectively,  neutron star models with $M =1.4 M_{\odot}$  and  $M =1.98 M_{\odot}$.

The $f$-mode instability may develop in the aftermath of a supernova explosion, when a new born neutron star may 
spin very rapidly. The most unstable $f$-mode is expected in the $l=m=3$ and 4 multipoles, 
which  have a relatively large instability window  with respect to the $l=m=2$ case.
  Considering various cases in which a neutron star enters the 
instability window at different rotation rates, we find that the gravitational-wave signal emitted during the instability 
may be detected by ET from a source located in the Virgo cluster, and as expected the gravitational signal is stronger for the more massive model with $N=0.62$. 
Actually for  this  model, the $l=m=3$ $f$-mode may be detected also by Advanced LIGO/Virgo. 
This result is valid for an $f$-mode which saturates at $E\simeq 10^{-6} M_{\odot}c^2$ and for a signal  integration time $\leq 1$yr. 
From the Virgo cluster we should expect about 30-60 supernova explosions per year, by assuming a rate of 2-3 events 
per century in our Galaxy. This means that if few of them leaves behind a very rapidly spinning proto-neutron star, 
we might be able to detect these events  from the gravitational radiation emitted during the $f$-mode instability.
 Stars with high mass and angular momentum are certainly more promising gravitational-wave sources. They may originate  
from roughy the $1\%$ of core collapses of progenitor stars with $M>10M_{\odot}$~\cite{2006ApJ...637..914W}. 
 The number of these potential sources may be even more promising if we note that in more massive stars the  
gravitational-wave signal may remain detectable for many years.

Another important result of our simulations is that the heat generated by shear viscosity during the $f$-mode saturation phase    
 prevents the star from entering the regime of mutual friction. In fact, the shear viscosity re-heating balances the neutrino cooling and
  leads to a nearly isothermal evolution. The star therefore  leaves the instability window 
 at lower rotation rates, and even a moderate change in the superfluid transition temperature does not change this result.

The magnetic torque as well as an unstable $r$-mode may accelerate the transition of the star through the instability 
window of the $f$-mode. This may limit the growth of the mode's amplitude and therefore the strain of the gravitational-wave radiation. 
Our results show that the magnetic field affects the $f$-mode instability when $B_{p} \gtrsim 10^{12}$~G, and its influence is more 
relevant in the $N=1$ neutron star model. For the $r$-mode we find that it must reach  
the maximum value expected from non-linear mode coupling studies in order to affect considerably 
the $f$-mode evolution. However, a definitive answer 
to this issue cannot be given as  the maximum $f$-mode amplitude  is still unknown.

In this study, we have not addressed  the 'classical' $l = m = 2$ $f$-mode instability, which is potentially the lowest order unstable mode driven by gravitational wave emission. As it was shown previously in \cite{2011PhRvL.107j1102G}, the prospects for a dynamical long-term evolution of a quadrupolar instability are rather pessimistic given the diminished size of the corresponding instability window. However, dropping the Cowling-approximation which we used here will certainly improve the situation for the $l = m = 2$ case since in full general relativity the critical angular velocity for the onset of the CFS-instability is shifted towards lower values \cite{2010PhRvD..81h4055Z}. On the other hand, including relativistic effects for the $r$-mode might have the opposite effect in the sense that relativistic perturbation theory hints towards a weakening, i.e.\ much larger growth times, of the instability when compared to Newtonian results \cite{Ruoff:2002lr}. This also works in favor of an enhanced detectability even if $f$- and $r$-mode are unstable at the same time.

So far we used polytropic equations of state for modelling relativistic neutron stars. A natural extension of this work should also incorporate a wide range of realistic equations of state. 
This most probably  will not change the main results of this work but for sure will lead to alterations in mode frequencies, growth times and gravitational wave detectability.  
At the same time, in more compact neutron star models one needs also to take into account the effect of direct Urca processes. These $\beta$-reaction processes are energetically favourable when 
the proton fraction is sufficiently large, $x_{\p} > 1/9$, \cite{1991PhRvL..66.2701L}, and they lead to 
a stronger bulk viscosity which may limit considerably the instability window of the $f$-mode~\cite{2012MNRAS.422.3327P}.

Another potentially significant damping mechanism neglected here is the formation of a crust and its importance on the $f$-mode instability. Additional dissipation will occur at a viscous boundary layer at the crust-core interface or more directly, the crust could break by a large amplitude $f$-mode. This would release a non-negligible amount of energy for dissipation at the fracture sites. Of course this mechanism depends on how fast a crust is formed after the creation of a hot, rapidly rotating proto-neutron star. Large amplitude oscillations might even delay the formation of the crust in the nascent neutron star. 
 
\section*{Acknowledgements}
AP acknowledge support from the German Science Foundation 
(DFG) via SFB/TR7, and from the European Commission Seventh Framework Programme (FP7/2007-2013) under grant agreement n$^{\rm o}$ 267251. 
EG acknowledge support from the German Science Foundation 
(DFG) via SFB/TR7.
DD acknowledges support from the German Science Foundation (DFG) via
SFB/TR7 and by the Bulgarian National Science Fund under Grant
DMU-03/6.
We would like to thank K. Glampedakis and L. Rezzolla for stimulating discussions and comments.

\appendix

\section{The evolution equations of the  $f$- and the $r$-mode instability} \label{sec:app}

In this Section, we derive the evolution equations for studying the evolution of the gravitational wave instability 
driven by two modes. Although these equations have been applied in Sec.~\ref{sec:RM} for studying the $f$- and $r$-modes 
the formalism is general. 
As in the single mode case the main  equations are~(\ref{eq:dEdt}) and~(\ref{eq:dJdt}), which now have to be extended to describe the 
instability of two modes.

The first two equations can be derived from the mode energy variation~(\ref{eq:dEdt})\begin{equation}
 \frac{d \alpha_{\x} }{dt} + \alpha_{\x} \frac{d \ln \tilde{E_{\x}}}{d\Omega} \frac{d\Omega}{dt} = -   \frac{2 \alpha_{\x}}{\tau_{\x}} \, ,  
\label{eq:evol1A}
\end{equation}
where the index $\x = \f,\ar$ identifies the quantities related to the $f$- and $r$-modes, respectively, 
and we have defined the energy of each x-mode as $E_{\x} = \alpha_{\x} \tilde E_{\x} \left( \Omega \right)$.
 
A third equation can be determined  from the angular momentum evolution equation~(\ref{eq:dJdt}), 
with the canonical angular momentum now given by 
\begin{equation}
J_c = \alpha_{\f} \tilde J^{\f} _c \left( \Omega \right) + \alpha_{\ar} \tilde J^{\ar} _c\left( \Omega \right) \, , 
\end{equation}
where $\tilde J^{\f} _c \left( \Omega \right)$ and $\tilde J^{\ar} _c \left( \Omega \right)$ describe the dependence 
of the canonical angular momentum of each mode on the angular velocity of the star. 
The other term that changes in equation~(\ref{eq:dJdt}) is the gravitational radiation torque, which now can be decomposed as
\begin{equation}
\frac{d J_{gw}}{ dt}  =  -   \frac{2 J_c^{\f} }{ \tau^{\f}_{\rm gw}}   -  \frac{2 J_c^{\ar} }{ \tau^{\ar}_{\rm gw}}   \,   , 
\end{equation}
where $\tau^{\f}_{\rm gw}$ and $\tau^{\ar}_{\rm gw}$ are the gravitational growth time of the $f$- and $r$-modes. 
Taking the time derivatives, the angular momentum conservation equation reads 
\begin{align}
& J_c^{\f} \frac{d \alpha_{\f} }{dt} + J_c^{\ar} \frac{d \alpha_{\ar} }{dt}  
+ \left( \frac{dJ_s}{d\Omega} + \alpha_{\f} \frac{d J^{\f}_c}{d\Omega} 
 + \alpha_{\ar} \frac{d J^{\ar}_c}{d\Omega}  \right) \frac{d\Omega}{dt} =
\nn \\ 
& -   \frac{2 J_c^{\f} }{ \tau^{\f}_{\rm gw}}   -  \frac{2 J_c^{\ar} }{ \tau^{\ar}_{\rm gw}}  
+  \frac{d J_{mag}}{dt} \, .  \label{eq:evol2A}
\end{align}

As final step, we can combine equations~(\ref{eq:evol1A}) and~(\ref{eq:evol2A}) and obtain the following system of evolution equations
\begin{align}
 \frac{d \alpha_{\x}}{dt} = &  - \frac{2 \alpha_{\x}}{\tau^{\x}_{\rm gw}} 
 - \frac{2 \alpha_{\x}}{\tau_{\rm v}^{\x}}  \frac{D_{x}}{D} 
 - \frac{2 \alpha_{\x} \al_{\y} }{\tau_{\rm v}^{\y}}  \frac{A_{\x} F_{y} } { D}  
 \nn \\  & 
 -  \alpha_{x} \frac{A_{\x}}{D}  \left(  \frac{d J_{s} } { d\Omega }   \right)^{-1}   \frac{d J_{mag}}{dt}  \, , \label{eq:daldtA} \\
D  \frac{d \Omega}{dt} = & 
\sum_{\x}    \frac{ 2 F_{\x} \alpha_{\x}}{\tau_{\rm v}^{\x}}  +  \left(  \frac{d J_{s} } { d\Omega }   \right)^{-1}   \frac{d J_{mag}}{dt} \,  , 
 \label{eq:dOmdtA}
\end{align}
where also $\y $ denotes the $f$- and $r$-mode quantities with the condition that $\x \neq \y$. 

In equations~(\ref{eq:daldtA}) and~(\ref{eq:dOmdtA}) we have defined the following functions:
\begin{align}
& A_{\x} = \frac{d \ln \tilde E_{\x}}{d\Omega} \, , \quad  \qquad  F_{\x} = \tilde J^{\x}_c \left( \frac{dJ_s}{d\Omega} \right) ^{-1} \, , \\
& P_{\x} = A_{\x} \cdot F_{\x}  \, ,  \qquad \quad \, \, Q_{\x} = \frac{d\tilde J_c^{\x}}{d\Omega} \left( \frac{dJ_s}{d\Omega} \right) ^{-1} \, , 
\end{align}
where 
\begin{align}
& D = 1 + \sum_{\x} \alpha_{\x}  \left( Q_{\x} - P_{\x} \right) \,  , \\ 
& D_{\x} = D + \alpha_{\x} P_{\x}  \, .
\end{align}
 
 Similar to the single mode instability, we may determine the evolution equations of a star in case one of the modes 
 saturates. When the y-mode saturates its amplitude is nearly constant,  i.e.   $d \alpha_{\y} / dt = 0$, 
 and equations~(\ref{eq:daldtA}) and~(\ref{eq:dOmdtA}) 
 now read
\begin{align}
 \frac{d \alpha_{\x}}{dt} = &  - \frac{2 \alpha_{\x}}{\tau^{\x}_{\rm gw}} 
 - \frac{2 \alpha_{\x} }{\tau_{\rm v}^{\x}}  \frac{ D_q }{D_{\y}} 
 + \frac{2 \alpha_{\x} \alpha_{\y} }{\tau_{\rm v}^{\y} }  \frac{A_{\x} F_{\y}  } {D_{\y}}  \nn  \\ 
  &  
    - \frac{   \alpha_{x}  A_{\x} }{D_{\y} }  \left(  \frac{d J_{s} } { d\Omega }   \right)^{-1}   \frac{d J_{mag}}{dt}  \, , \label{eq:daldtB} \\ 
D_{\y}  \frac{d \Omega}{dt} = & - \frac{ 2 F_{\y}  \al_{\y} }  {\tau^{\y} _{\rm gw} } 
 +    \frac{ 2 F_{\x} \alpha_{\x}}{\tau_{\rm v}^{\x}}  +  \left(  \frac{d J_{s} } { d\Omega }   \right)^{-1}   \frac{d J_{mag}}{dt} \, ,  
\end{align}
 where $D_{q} = 1 + \sum_{\x} \al_{\x} Q_{\x} $.

When both modes have reached their respective saturation amplitudes we approximately have that 
$d \alpha_{\f} / dt = d \alpha_{\ar} / dt = 0$. As a result the star rotation evolves according to the following equation
\begin{equation}
 D_{q} \frac{d \Omega}{dt} = 
-  2 \sum_{\x}    \frac{ F_{\x} \alpha_{\x}}{\tau_{\rm gw}^{\x}}  +  \left(  \frac{d J_{s} } { d\Omega }   \right)^{-1}   \frac{d J_{mag}}{dt} \, ,  
\end{equation}
where in a low magnetised star  both $\tau^{\f}_{\rm gw}$ and $\tau^{\ar}_{\rm gw}$ govern the spindown-rate of the star.

 \nocite*

\end{document}